\documentclass[oribibl, 11pt]{llncs}

\usepackage{amsmath}
\usepackage{amssymb}
\usepackage{graphicx}
\usepackage{subfigure}
\usepackage{algorithm}
\usepackage{algpseudocode}
\usepackage{color}
\usepackage{hyperref}

\usepackage{fancyhdr}
\newcommand{\removed}[1]{}

\newcommand{\rem}{\mathbf}

\newcommand{\cg}{\mathcal{G}}

\newcommand{\calt}{\mathcal{T}}

\newcommand{\ra}{\rightarrow}

\newcommand{\lra}{\Leftrightarrow}

\newcommand{\bs}{\backslash}

\newcommand{\dprime}{\prime\prime}

\DeclareMathOperator{\lcm}{\textup{lcm}}

\renewcommand{\P}{\textup{P}}
\newcommand{\E}{\textup{E}}
\newcommand{\Var}{\textup{Var}}

\newcommand{\OPT}{\text{OPT}}

\definecolor{DarkRed}{RGB}{182,11,1}

\title{An Introduction to Temporal Graphs: An Algorithmic Perspective\thanks{Supported in part by the project ``Foundations of Dynamic Distributed Computing Systems'' (\textsf{FOCUS}) which is implemented under the ``ARISTEIA'' Action of the  Operational Programme ``Education and Lifelong Learning'' and is co-funded by the European Union (European Social Fund) and Greek National Resources.}}

\author{Othon Michail}
\institute{Computer Technology Institute \& Press ``Diophantus'' (CTI),\\N. Kazantzaki Str., Patras University Campus,\\ Rio, P.O. Box 1382, 26504,\\Patras, Greece\\
Email:\email{ michailo@cti.gr}\\ Phone: +30 2610 960300}

\begin{document}

\maketitle

\begin{abstract}
A \emph{temporal graph} is, informally speaking, a graph that changes with time.  When time is discrete and only the relationships between the participating entities may change and not the entities themselves, a temporal graph may be viewed as a sequence $G_1,G_2\ldots,G_l$ of static graphs over the same (static) set of nodes $V$. Though static graphs have been extensively studied, for their temporal generalization we are still far from having a concrete set of structural and algorithmic principles. Recent research shows that many graph properties and problems become radically different and usually substantially more difficult when an extra time dimension in added to them. Moreover, there is already a rich and rapidly growing set of modern systems and applications that can be naturally modeled and studied via temporal graphs. This, further motivates the need for the development of a temporal extension of graph theory. We survey here recent results on temporal graphs and temporal graph problems that have appeared in the Computer Science community.
\end{abstract}

\section{Introduction}
\label{sec:intro}

The conception and development of graph theory is probably one of the most important achievements of mathematics and combinatorics of the last few centuries. Its applications are inexhaustible and ubiquitous. Almost every scientific domain, from mathematics and computer science to chemistry and biology, is a natural source of problems of outstanding importance that can be naturally modeled and studied by graphs. The 1736 paper of Euler on the Seven Bridges of K{\" o}nigsberg problem is regarded as the first formal treatment of a graph-theoretic problem. Till then, graph theory has found applications in electrical networks, theoretical chemistry, social network analysis, computer networks (like the Internet) and distributed systems, to name a few, and has also revealed some of the most outstanding problems of modern mathematics like the four color theorem and the traveling salesman problem.

Graphs simply represent a set of objects and a set of pairwise relations between them. It is very common, and shows up in many applications, the pairwise relations to come with some additional information. For example, in a graph representing a set of cities and the available roads from each city to the others, the additional information of an edge $(C_1,C_2)$ could be the average time it takes to drive from city $C_1$ to city $C_2$. In a graph representing bonding between atoms in a molecule, edges could also have an additional bond order or bond strength information. Such applications can be modeled by weighted or, more generally, by labeled graphs, in which edges (and in some cases also nodes) are assigned values from some domain, like the set of natural numbers. An example of a classical, very rich, and well-studied area of labeled graphs is the area of graph coloring \cite{MR02}.

\emph{Temporal graphs} (also known as \emph{dynamic}, \emph{evolving} \cite{Fe04}, or \emph{time-varying} \cite{FMS09,CFQS12} graphs) can be informally described as \emph{graphs that change with time}. In terms of modeling, they can be thought of as a special case of labeled graphs, where labels capture some measure of time. Inversely, it is also true that any property of a graph labeled from a discrete set of labels corresponds to some temporal property if interpreted appropriately. For example, a proper edge-coloring, i.e. a coloring of the edges in which no two adjacent edges share a common color, corresponds to a temporal graph in which no two adjacent edges share a common time-label, i.e. no two adjacent edges ever appear at the same time. Still, the time notion and the rich domain of modern applications motivating its incorporation to graphs, gives rise to a brand new set of challenging, important, and practical problems that could not have been observed from the more abstract perspective of labeled graphs.

Though the formal treatment of temporal graphs is still in its infancy, there is already a huge identified set of applications and research domains that motivate it and that could benefit from the development of a concrete set of results, tools, and techniques for temporal graphs. A great variety of both modern and traditional networks such as information and communication networks, social networks, transportation networks, and several physical systems can be naturally modeled as temporal graphs. In fact, this is true for almost any network with a dynamic topology. Most modern communication networks, such as mobile ad-hoc, sensor, peer-to-peer, opportunistic, and delay-tolerant networks, are inherently dynamic. In social networks, the topology usually represents the social connections between a group of individuals and it changes as the social relationships between the individuals are updated, or as individuals leave or enter the group. In a transportation network, there is usually some fixed network of routes and a set of transportation units moving over these routes and dynamicity refers to the change of the positions of the transportation units in the network as time passes. Physical systems of interest may include several systems of interacting particles or molecules reacting in a well-mixed solution. Temporal relationships and temporal ordering of events are also present in the study of epidemics, where a group of individuals (or computing entities) come into contact with each other and we want to study the spread of an infectious disease (or a computer virus) in the population.

A very rich motivating domain is that of distributed computing systems that are inherently dynamic. The growing interest in such systems has been mainly driven by the advent of low-cost wireless communication devices and the development of efficient wireless communication protocols. Apart from the huge amount of work that has been devoted to applications, there is also a steadily growing concrete set of foundational work. A notable set of works has studied (distributed) computation in \emph{worst-case} dynamic networks in which the topology may change arbitrarily from round to round subject to some constraints that allow for bounded end-to-end communication~\cite{OW05,KLO10,MCS14,MCS13b,DPRS13,APRU12}. Population protocols \cite{AADFP06} and variants \cite{MCS11-2,MS14b} are collections of finite-state agents that move passively, according to the dynamicity of the environment, and interact in pairs when they come close to each other. The goal is typically for the population to compute (i.e. agree on) something useful or construct a desired network or structure in such an adversarial setting. Another interesting direction assumes that the dynamicity of the network is a result of randomness (this is also the case sometimes in population protocols). Here the interest is on determining ``good'' properties of the dynamic network that hold with high probability (abbreviated w.h.p. and meaning with probability at least $1-1/n^c$ for some constant $c\geq 1$), such as small (temporal) diameter, and on designing protocols for distributed tasks \cite{CFTE08,AKL08}. In all the above subjects, there is always some sort of underlying temporal graph either assumed or implied. For introductory texts on the above lines of research in dynamic distributed networks the reader is referred to \cite{CFQS12,MCS11,Sc02,KO11}.

Though static graphs \footnote{In this article, we use ``static'' to refer to classical graphs. This is plausible as the opposite of ``dynamic'' that is also commonly used for temporal graphs. In any case, the terminology is still very far from being standard.} have been extensively studied, for their temporal generalization we are still far from having a concrete set of structural and algorithmic principles. Additionally, it is not yet clear how is the complexity of combinatorial optimization problems affected by introducing to them a notion of time. In an early but serious attempt to answer this question, Orlin \cite{Or81} observed that many dynamic languages derived from $\rem{NP}$-complete languages can be shown to be $\rem{PSPACE}$-complete. Among the other few things that we do know, is that the max-flow min-cut theorem holds with unit capacities for time-respecting paths \cite{Be96}. Additionally, Kempe \emph{et al.} \cite{KKK00} proved that, in temporal graphs, the classical formulation of Menger's theorem is violated and the computation of the number of node-disjoint $s$-$z$ paths becomes $\rem{NP}$-complete. A reformulation of Menger's theorem which is valid for all temporal graphs was recently achieved in \cite{MMCS13}. These results are discussed in Section \ref{sec:menger}. Recently, building on the distributed online dynamic network model of \cite{KLO10}, Dutta \emph{et al.} \cite{DPRS13}, among other things, presented \emph{offline centralized algorithms} for the $k$-\emph{token dissemination} problem. In $k$-token dissemination, there are $k$ distinct pieces of information (tokens) that are initially present in some distributed processes and the problem is to disseminate all the $k$ tokens to all the processes in the dynamic network, under the constraint that one token can go through an edge per round. These results, motivated by distributed computing systems, are presented in Section \ref{sec:dissemination}.

Another important problem is that of \emph{designing} an efficient temporal graph given some requirements that the graph should meet. This problem was recently studied in \cite{MMCS13}, where the authors introduced several interesting cost minimization parameters for optimal temporal network design. One of the parameters is the \emph{temporality} of a graph $G$, in which the goal is to create a temporal version of $G$ minimizing the maximum number of labels of an edge, and the other is the \emph{temporal cost} of $G$, in which the goal is to minimize the total number of labels used. Optimization of these parameters is performed subject to some \emph{connectivity constraint}. They proved several upper and lower bounds for the temporality of some very basic graph families such as rings, directed acyclic graphs, and trees, as well as a trade-off between the temporality and the maximum label of rings. Furthermore, they gave a \emph{generic method} for computing a lower bound of the temporality of an arbitrary graph with respect to (abbreviated w.r.t.) the constraint of preserving a time-respecting analogue of every simple path of $G$. Finally, they proved that computing the temporal cost w.r.t. the constraint of preserving at least one time-respecting path from $u$ to $v$ whenever $v$ is reachable from $u$ in $G$, is $\rem{APX}$-hard. Most of these results are discussed in Section \ref{sec:design}.

Other recent papers have focused on understanding the complexity and providing algorithms for temporal versions of classical graph problems. For example, the authors of \cite{MS14} considered temporal analogues of \emph{traveling salesman problems} (TSP) in temporal graphs, and in the way also introduced and studied temporal versions of other fundamental problems like {\sc Maximum Matching}, {\sc Path Packing}, {\sc Max-TSP}, and {\sc Minimum Cycle Cover}. One such version of TSP is the problem of exploring the nodes of a temporal graph as soon as possible. In contrast to the positive results known for the static case strong inapproximability results can be proved for the dynamic case \cite{MS14,EHK15}. Still, there is room for positive results for interesting special cases \cite{EHK15}. Another such problem is the {\sc Temporal Traveling Salesman Problem with Costs One and Two} (abbreviated TTSP(1,2)), a temporal analogue of TSP(1,2), in which the temporal graph is a complete weighted graph with edge-costs from $\{1,2\}$ and the cost of an edge may vary from instance to instance \cite{MS14}. The goal is to find a minimum cost temporal TSP tour. Several \emph{polynomial-time approximation algorithms} have been proved for TTSP(1,2) \cite{MS14}. The best approximation is $(1.7+\varepsilon)$ for the generic TTSP(1,2) and $(13/8+\varepsilon)$ for its interesting special case in which the lifetime of the temporal graph is restricted to $n$. These and related results are presented in Section \ref{sec:matching}.

Additionally, there are works that have considered \emph{random temporal graphs}, in which the labels are chosen according to some probability distribution. We give a brief introduction to such models in Section \ref{sec:random}. Moreover, Section \ref{sec:model} provides all necessary preliminaries and definitions and also a first discussion on temporal paths and Section \ref{sec:linear} discusses a temporal graph model in which the availability times of the edges are provided by a set of linear functions.

As is always the case, not all interesting results and material could fit in a single document. We list here some of them. Holme and Saram{\" a}ki \cite{HS12} give an extensive overview of the literature related to temporal networks from a diverge range of scientific domains. Harary and Gupta \cite{HG97} discuss applications of temporal graphs and highlight the great importance of a systematic treatment of the subject. Kostakos \cite{Ko09} uses temporal graphs to represent real datasets, shows how to derive various node metrics like average temporal proximity, average geodesic proximity and temporal availability, and also gives a static representation of a temporal graph (similar to the \emph{static expansion} that we discuss in Section \ref{sec:model}). Avin \emph{et al.} \cite{AKL08} studied the cover time of a simple random walk on Markovian dynamic graphs and proved that, in great contrast to being always polynomial in static graphs, it is exponential in some dynamic graphs. Clementi \emph{et al.} \cite{CFTE08} studied the flooding time (also known as information dissemination; see a similar problem discussed in Section \ref{sec:dissemination}) in the following type of edge-markovian dynamic graphs: if an edge exists at time $t$ then, at time $t + 1$, it disappears with probability $q$, and if instead the edge does not exist at time $t$, then it appears at time $t + 1$ with probability $p$. There are also several papers that have focused on temporal graphs in which every instance of the graph is drawn independently at random according to some distribution \cite{CPMS07,HHL88,Pi87,KK02} (the last three did it in the context of dynamic gossip-based mechanisms), e.g. according to $\cg(n,p)$. A model related to random temporal graphs, is the \emph{random phone-call model}, in which each node, at each step, can communicate with a random neighbour \cite{DGH87,KSSV00}. Other authors \cite{XFJ03,FT98} have assumed that an edge may be available for a whole time-interval $[t_1,t_2]$ or several such intervals and not just for discrete moments. Aaron \emph{et al.} \cite{AKM14} studied the {\sc Dynamic Map Visitation} problem, in which a team of agents must visit a collection of critical locations as quickly as possible in a dynamic environment. Kontogiannis \emph{et al.} \cite{KMP15}, among other things, presented oracles for providing time-dependent min-cost route plans and conducted their experimental evaluation on a data set of the city of Berlin. 

\section{Modeling and Basic Properties}
\label{sec:model}

When time is assumed to be discrete, a temporal graph (or digraph) is just a static graph (or digraph) $G=(V,E)$ with every edge $e\in E$ labeled with zero or more natural numbers. The labels of an edge may be viewed as the times at which the the edge is \emph{available}. For example, an edge with no labels is never available while, on the other hand, an edge with labels all the even natural numbers is available every even time. Labels could correspond to seconds, days, years, or could even correspond to some artificial discrete measure of time under consideration. 

There are several ways of modeling formally discrete temporal graphs. One is to consider an underlying static graph $G=(V,E)$ together with a labeling $\lambda:E\rightarrow 2^\bbbn$ of $G$ assigning to every edge of $G$ a (possibly empty) set of natural numbers, called \emph{labels}. Then the temporal graph of $G$ with respect to $\lambda$ is denoted by $\lambda(G)$. This notation is particularly useful when one wants to explicitly refer to and study properties of the labels of the temporal graph. For example, the multiset of all labels of $\lambda(G)$ can be denoted by $\lambda(E)$, their cardinality is defined as $|\lambda|=\sum_{e\in E} |\lambda(e)|$, and the maximum and minimum label assigned to the whole temporal graph as $\lambda_{\max}= \max\{l\in \lambda(E)\}$ and $\lambda_{\min}= \min\{l\in \lambda(E)\}$, respectively. Moreover, we define the \emph{age} (or \emph{lifetime}) of a temporal graph $\lambda(G)$ as $\alpha(\lambda)= \lambda_{\max}-\lambda_{\min}+1$ (or simply $\alpha$ when clear from context). Note that in case $\lambda_{\min}=1$ then we have $\alpha(\lambda)=\lambda_{\max}$.

Another, often convenient, notation of a temporal graph $D$ is as an ordered pair of disjoint sets $(V,A)$ such that $A\subseteq \binom{V}{2}\times\bbbn$ in case of a graph and with $\binom{V}{2}$ replaced by $V^2\bs \{(u,u):u\in V\}$ in case of a digraph. The set $A$ is called the set of \emph{time-edges}. $A$ can also be used to refer to the structure of the temporal graph at a particular time. In particular, $A(t)= \{e: (e,t)\in A\}$ is the (possibly empty) set of all edges that appear in the temporal graph at time $t$. In turn, $A(t)$ can be used to define a snapshot of the temporal graph $D$ at time $t$, which is usually called the $t$-\emph{th instance of $D$}, and is the static graph $D(t)= (V,A(t))$. So, it becomes evident that a temporal graph may also be viewed as a \emph{sequence of static graphs} $(G_1,G_2,\ldots,G_{\lambda_{\max}})$.

Finally, it is typically very useful to expand in time the whole temporal graph and obtain an equivalent static graph without losing any information. The reason for doing this is mainly because static graphs are much better understood and there is a rich set of well established tools and techniques for them. So, a common approach to solve a problem concerning temporal graphs is to first express the given temporal graph as a static graph and then try to apply or adjust one of the existing tools that works on static graphs. Formally, the \emph{static expansion} of a temporal graph $D=(V,A)$ is a DAG $H=(S,E)$ defined as follows. If $V=\{u_1,u_2,\ldots,u_n\}$ then $S= \{u_{ij}: \lambda_{\min}-1\leq i\leq \lambda_{\max},1\leq j\leq n\}$ and $E=\{(u_{(i-1)j},u_{ij^\prime}):\lambda_{\min}\leq i\leq \lambda_{\max}$ and $j=j^\prime$ or $(u_j,u_{j^\prime})\in A(i)\}$. In words, for every discrete moment we create a copy of $V$ representing the instance of the nodes at that time (called \emph{time-nodes}). We may imagine the moments as levels or rows from top to bottom, every level containing a copy of $V$. Then we add outgoing edges from time-nodes of one level only to time-nodes of the level below it. In particular, we connect a time-node $u_{(i-1)j}$ to its own subsequent copy $u_{ij}$ and to every time-node $u_{ij^\prime}$ s.t. $(u_j,u_{j^\prime})$ is an edge of the temporal graph at time $i$. Observe that the above construction includes all possible vertical edges from a node to its own subsequent instance. These edges express the fact that nodes are usually not oblivious and can preserve their on history in time (modeled like propagating information to themselves). Nevertheless, depending on the application, these edges may some times be omitted. 

\subsection{Journeys}
\label{subsec:journeys}

As is the case in static graphs, the notion of a \emph{path} is one of the most central notions of a temporal graph, however it has to be redefined to take time into account. A \emph{temporal} (or \emph{time-respecting}) \emph{walk} $W$ of a temporal graph $D=(V,A)$ is an alternating sequence of nodes and times $(u_1,t_1,u_2,t_2,\ldots,u_{k-1},t_{k-1},u_k)$ where $(u_iu_{i+1},t_i)\in A$, for all $1\leq i\leq k-1$, and $t_i<t_{i+1}$, for all $1\leq i\leq k-2$. We call $t_{k-1}-t_1+1$ the \emph{duration} (or \emph{temporal length}) of the walk $W$, $t_1$ its \emph{departure time} and $t_{k-1}$ its \emph{arrival time}. A \emph{journey} (or \emph{temporal/time-respecting path}) $J$ is a temporal walk with pairwise distinct nodes. In words, a journey of $D$ is a path of the underlying static graph of $D$ that uses strictly increasing edge-labels. A $u$-$v$ journey $J$ is called \emph{foremost from time $t\in\bbbn$} if it departs after time $t$ and its arrival time is minimized. The \emph{temporal distance} from a node $u$ at time $t$ to a node $v$ is defined as the duration of a foremost $u$-$v$ journey from time $t$. We say that a temporal graph $D=(V,A)$ has \emph{temporal (or dynamic) diameter} $d$, if $d$ is the minimum integer for which it holds that the temporal distance from every time-node $(u,t)\in V\times \{0,1,\ldots,\alpha-d\}$ to every node $v\in V$ is at most $d$.

A nice property of foremost journeys is that they can be computed efficiently. In particular there is an algorithm that, given a source node $s\in V$ and a time $t_{start}$, computes for all $w\in V\bs\{s\}$ a foremost $s$-$w$ journey from time $t_{start}$ \cite{MMCS13,MMS15}. The running time of the algorithm is $O(n\alpha^3(\lambda)+|\lambda|)$, where $n$ here and throughout this article denotes the number of nodes of the temporal graph. It is worth mentioning that this algorithm takes as input the whole temporal graph $D$. Such algorithms are known as \emph{offline} algorithms in contrast to \emph{online} algorithms to which the temporal graph is revealed on the fly. The algorithm is essentially a temporal translation of the breadth-first search (BFS) algorithm (see e.g. \cite{CLRS01} page 531) with path length replaced by path arrival time. For every time $t$, the algorithm picks one after the other all nodes that have been already reached (initially only the source node $s$) and inspects all edges that are incident to that node at time $t$. If a time-edge $(e,t)$ leads to a node $w$ that has not yet been reached, then $(e,t)$ is picked as an edge of a foremost journey from the source to $w$. This greedy algorithm is correct for the same reason that the BFS algorithm is correct. An immediate way to see this is by considering the static expansion of the temporal graph. The algorithm begins from the upper copy (i.e. at level 0) of the source in the static expansion and essentially executes the following slight variation of BFS: at step $i+1$, given the set $R$ of already reached nodes at level $i$, the algorithm first follows all vertical edges leaving $R$ in order to reach in one step the $(i+1)$-th copy of each node in $R$, and then inspects all diagonal edges leaving $R$ to discover new reachabilities. The algorithm outputs as a foremost journey to a node $u$, the directed path of time-edges by which it first reached the column of $u$ (vertical edges are interpreted as waiting on the corresponding node). The above algorithm computes a shortest path to each column of the static expansion. Correctness follows from the fact that shortest paths to columns are equivalent to foremost journeys to the nodes corresponding to the columns.

\section{Connectivity and Menger's Theorem}
\label{sec:menger}

Assume that we are given a static graph $G$ and a source node $s$ and a sink node $z$ of $G$. \footnote{The sink is usually denoted by $t$ in the literature. We use $z$ instead as we reserve $t$ to refer to time moments.} Two paths from $s$ to $z$ are called node-disjoint if they have only the nodes $s$ and $z$ in common. \emph{Menger's theorem} \cite{Me27}, which is the analogue of the max-flow min-cut theorem for undirected graphs, is one of the most basic theorems in the theory of graph connectivity. It states that \emph{the maximum number of node-disjoint $s$-$z$ paths is equal to the minimum number of nodes that must be removed in order to separate $s$ from $z$} (see also \cite{Bo98} page 75). 

\begin{figure}[!hbtp]
   \centering{
        \includegraphics[width=0.65\textwidth]{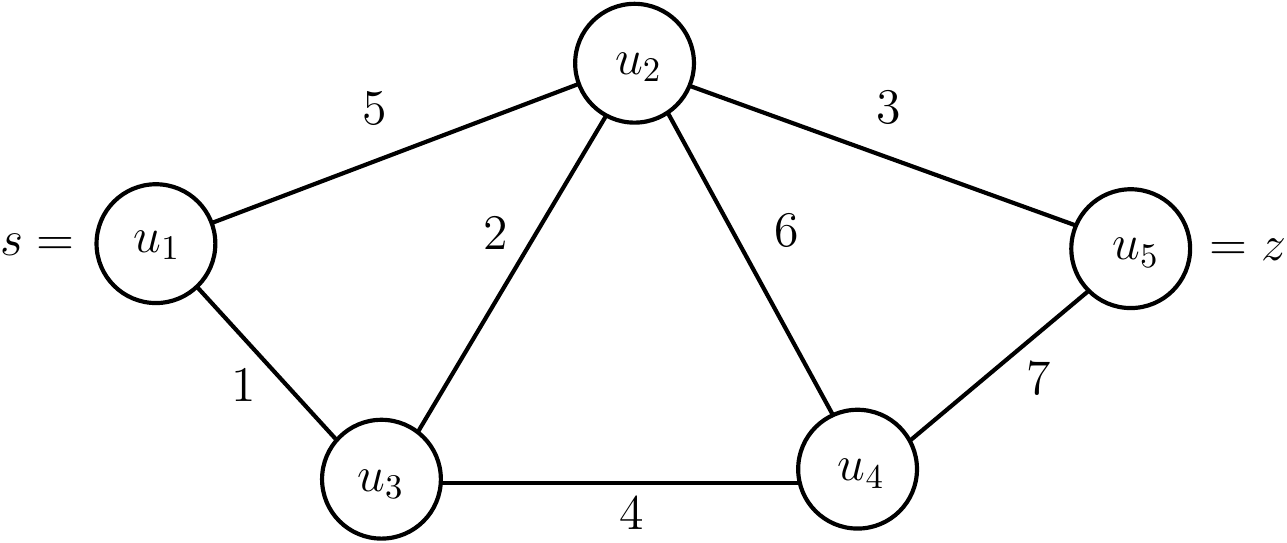}
        }
   \caption{A counterexample of Menger's theorem for temporal graphs (adopted from \cite{KKK00}). Each edge has a single time-label indicating its availability time.} \label{fig:ber}
\end{figure}

It was first observed in \cite{Be96} and then further studied in \cite{KKK00} that this fundamental theorem of static graphs, is violated in temporal graphs if we keep its original formulation and only require it to hold for journeys instead of paths. In fact, the violation holds even for a very special case of temporal graphs, those in which every edge has at most one label, which are known as \emph{single-labeled temporal graphs} (as opposed to the more general \emph{multi-labeled} temporal graphs that we have discussed so far). Even in such temporal graphs, the maximum number of node-disjoint journeys from $s$ to $z$ can be strictly less than the minimum number of nodes whose deletion leaves no $s$-$z$ journey. For a simple example, observe in Figure \ref{fig:ber} that there are no two node-disjoint journeys from $s$ to $z$ but after deleting any one node (other than $s$ or $z$) there still remains a $s$-$z$ journey. To see this, notice that every journey has to visit at least two of the inner-nodes $u_2,u_3,u_4$. If $u_2$ is one of them, then a vertical obstacle is introduced which cannot be avoided by any other journey. If $u_2$ is not, then the only disjoint path remaining is $(s,u_2,z)$ which is not a journey. On the other hand, any set of two inner vertices has a $s$-$z$ journey going through them implying that any $s$-$z$ separator must have size at least 2. As shown in \cite{KKK00}, this construction can be generalized to a single-labeled graph with $2k-1$ inner nodes in which: (i) every $s$-$z$ journey visits at least $k$ of these nodes, ensuring again that there are no two node-disjoint $s$-$z$ journeys and (ii) there is a journey through any set of $k$ inner nodes, ensuring that every $s$-$z$ separator must have size at least $k$.

On the positive side, the violation does not hold if we replace node-disjointness by edge-disjointness and node removals by edge removals. In particular, it was proved in \cite{Be96} that for single-labeled temporal graphs,  the maximum number of edge-disjoint journeys from $s$ to $z$ is equal to the minimum number of edges whose deletion leaves no $s$-$z$ journey, that is, that the max-flow min-cut theorem of static graphs holds with unit capacities for journeys in single-labeled temporal graphs. The construction (which we adopt from \cite{KKK00}) is simply an \emph{ad-hoc} static expansion for the special case of single-labeled temporal graphs. Let $G = (V,E)$ be the underlying graph of an undirected single-labeled temporal graph. We construct a labeled directed graph $G^{\prime} = (V^{\prime}, E^{\prime})$ as follows. for every $\{u, v\}\in E$ we add in $G^{\prime}$ two new nodes $x$ and $y$ and the directed edges $(u, x)$, $(v, x)$, $(y, u)$, $(y, v)$, $(x, y)$. Then we relax all labels required so that there is sufficient ``room'' (w.r.t. time) to introduce (by labeling the new edges) both a $(u,x,y,v)$ journey and a $(v,x,y,u)$ journey. The goal is to be able to both move by a journey from $u$ to $v$ and from $v$ to $u$ in $G^{\prime}$. An easy way to do this is the following: if $t$ is the label of $\{u, v\}$, then we can label $(u,x),(x,y),(y,v)$ by $(t.1,t.2,t.3)$, where $t.1<t.2<t.3$, and similarly for $(v,x),(x,y),(y,u)$. Then we construct a static directed graph $G^{\dprime} = (V^{\dprime}, E^{\dprime})$ as follows: For every $u\in V$ let $y_1,y_2,\ldots,y_i,\ldots$ be its incoming edges and $x_1, x_2,\ldots, x_j,\ldots$ its outgoing edges. We want to preserve only the time-respecting $y, u, x$ traversals. To this end, for each one of the $(y_i, u)$ edges we introduce a node $w_i$ and the edge $(y_i, w_i)$ and for each one of the $(u, x_i)$ edges we introduce a node $v_j$ and the edge $(v_j, x_j)$ and we delete node $u$. Finally, we introduce the edge $(w_i, v_j)$ iff $(y_i, u), (u, x_j)$ is time-respecting. This reduction preserves edge-disjointness and sizes of edge separators and if we add a super-source and a super-sink to $G^{\dprime}$ the max-flow min-cut theorem for static directed graphs yields the aforementioned result. Another interesting thing is that reachability in $G$ under journeys corresponds to (path) reachability in $G^{\dprime}$ so that we can use BFS on $G^{\dprime}$ to answer questions about foremost journeys in $G$, as we did with the static expansion in Section \ref{subsec:journeys}.

\begin{figure}[!hbtp]
   \centering{
        \includegraphics[width=0.8\textwidth]{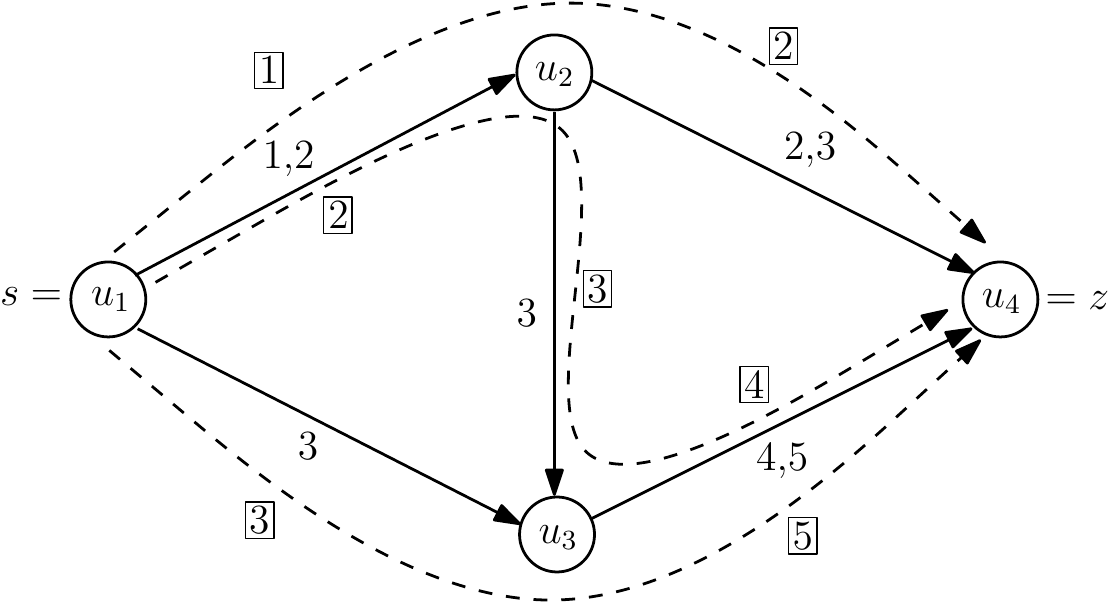}
        }
   \caption{An example of a temporal graph. The dashed curves highlight the directions of three out-disjoint journeys from $s$ to $z$. The labels used by each of these journeys are indicated by the labels that are enclosed in boxes.} \label{fig:menger-example}
\end{figure}

Fortunately, the above important negative result concerning Menger's theorem has a turnaround. In particular, it was proved in \cite{MMCS13} that if one reformulates Menger's theorem in a way that takes time into account then a very natural temporal analogue of Menger's theorem is obtained, which is valid for all (multi-labeled) temporal networks. The idea is to replace in the original formulation node-disjointness by \emph{node departure time disjointness} (or \emph{out-disjointness}) and node removals by \emph{node departure times removals}. When we say that we remove \emph{node departure time} $(u,t)$ we mean that we remove \emph{all edges leaving $u$ at time $t$}, i.e. we remove label $t$ from all $(u,v)$ edges (for all $v\in V$). So, when we ask ``how many node departure times are needed to separate two nodes $s$ and $z$?'' we mean how many node departure times must be selected so that after the removal of all the corresponding time-edges the resulting temporal graph has no $s$-$z$ journey (note that this is a different question from how many time-edges must be removed and, in fact, the latter question does not result in a Menger's analogue). Two journeys are called \emph{out-disjoint} if they never leave from the same node at the same time (see Figure \ref{fig:menger-example} for an example).

\begin{theorem} [Menger's Temporal Analogue \cite{MMCS13}] \label{the:dmeng}
Take any temporal graph $\lambda(G)$, where $G=(V,E)$, with two distinguished nodes $s$ and $z$. The maximum number of out-disjoint journeys from $s$ to $z$ is equal to the minimum number of node departure times needed to separate $s$ from $z$.  
\end{theorem}

\begin{figure}[!hbtp]
   \centering{
        \subfigure[]{
        \includegraphics[width=0.362\textwidth]{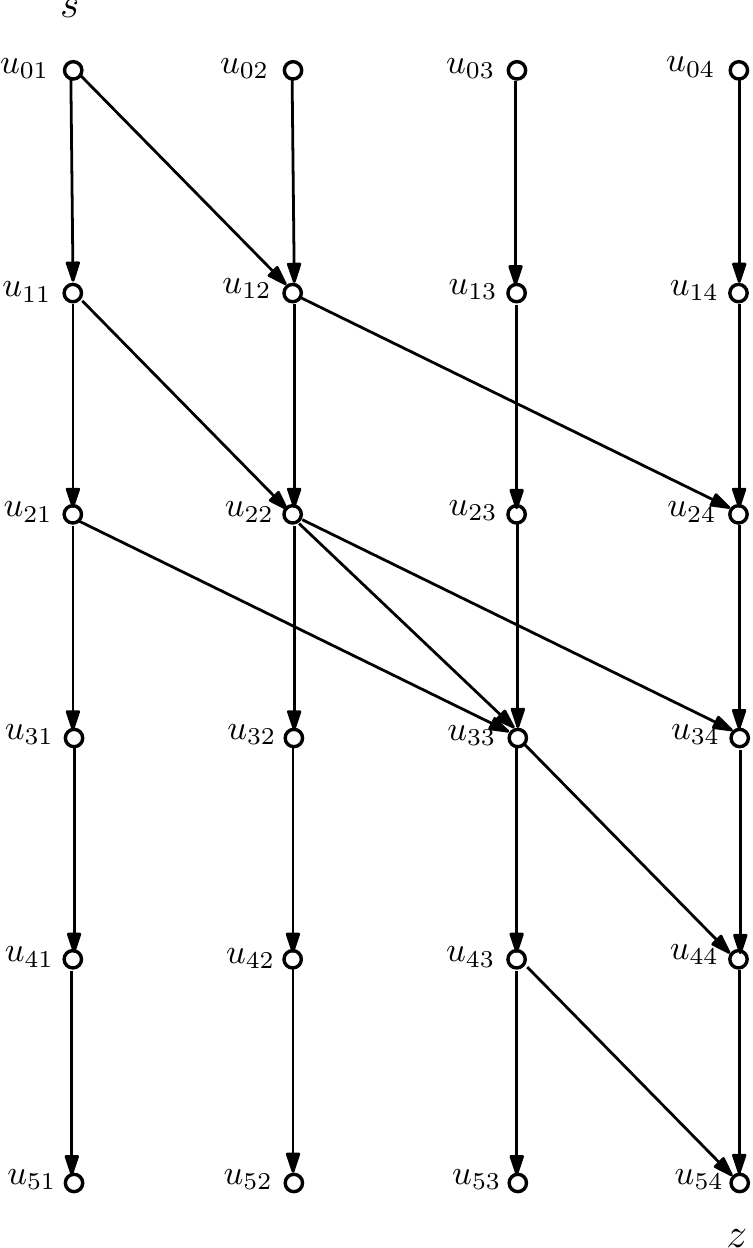}
        \label{fig:shape1}}
	\hspace{1cm}
        \subfigure[]{
        \includegraphics[width=0.48\textwidth]{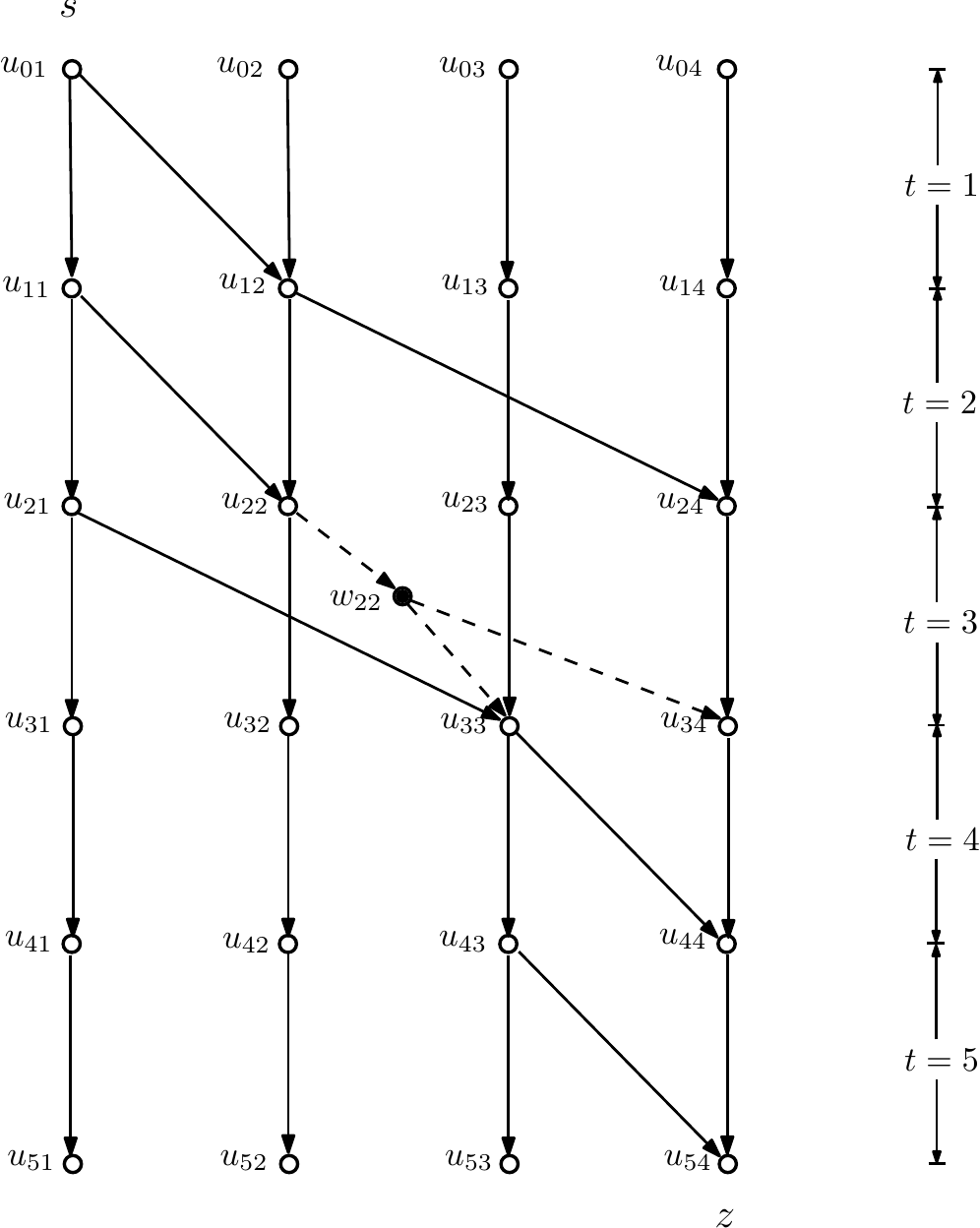}
        \label{fig:shape2}}
	\hspace{1cm}
        \subfigure[]{
        \includegraphics[width=0.362\textwidth]{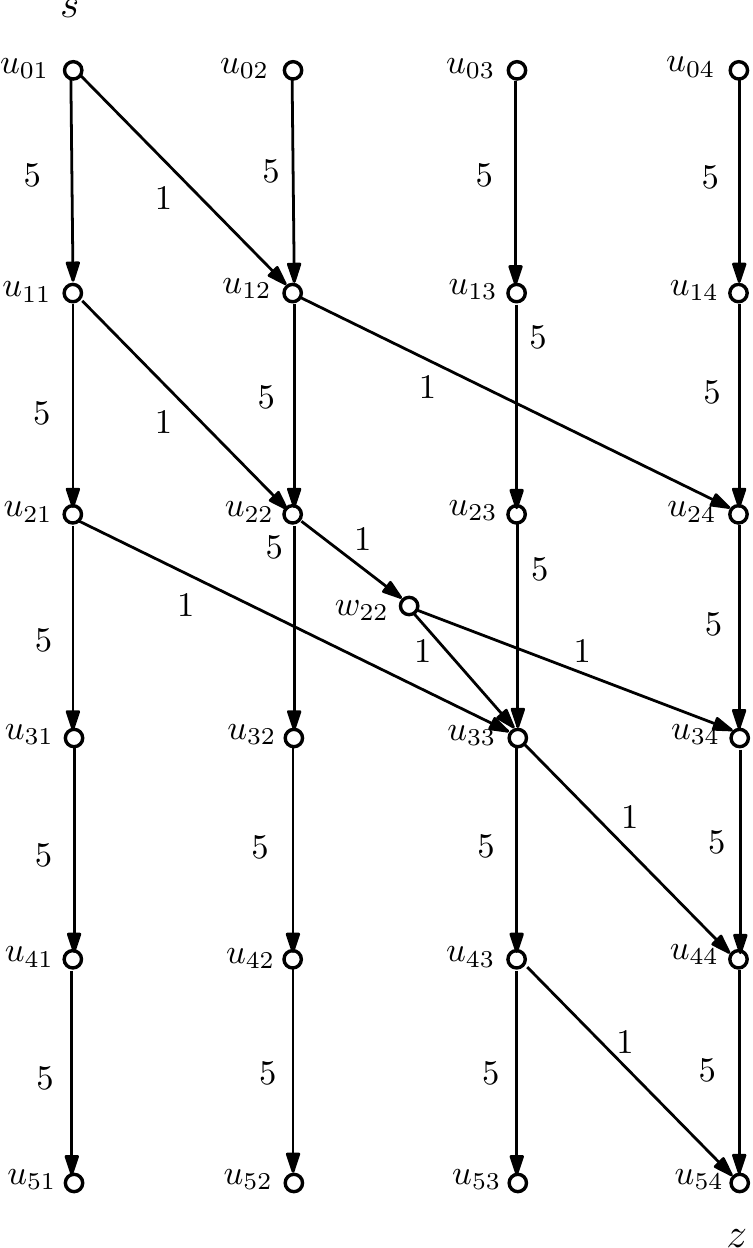}
        \label{fig:shape3}}
	\hspace{1cm}
        \subfigure[]{
        \includegraphics[width=0.48\textwidth]{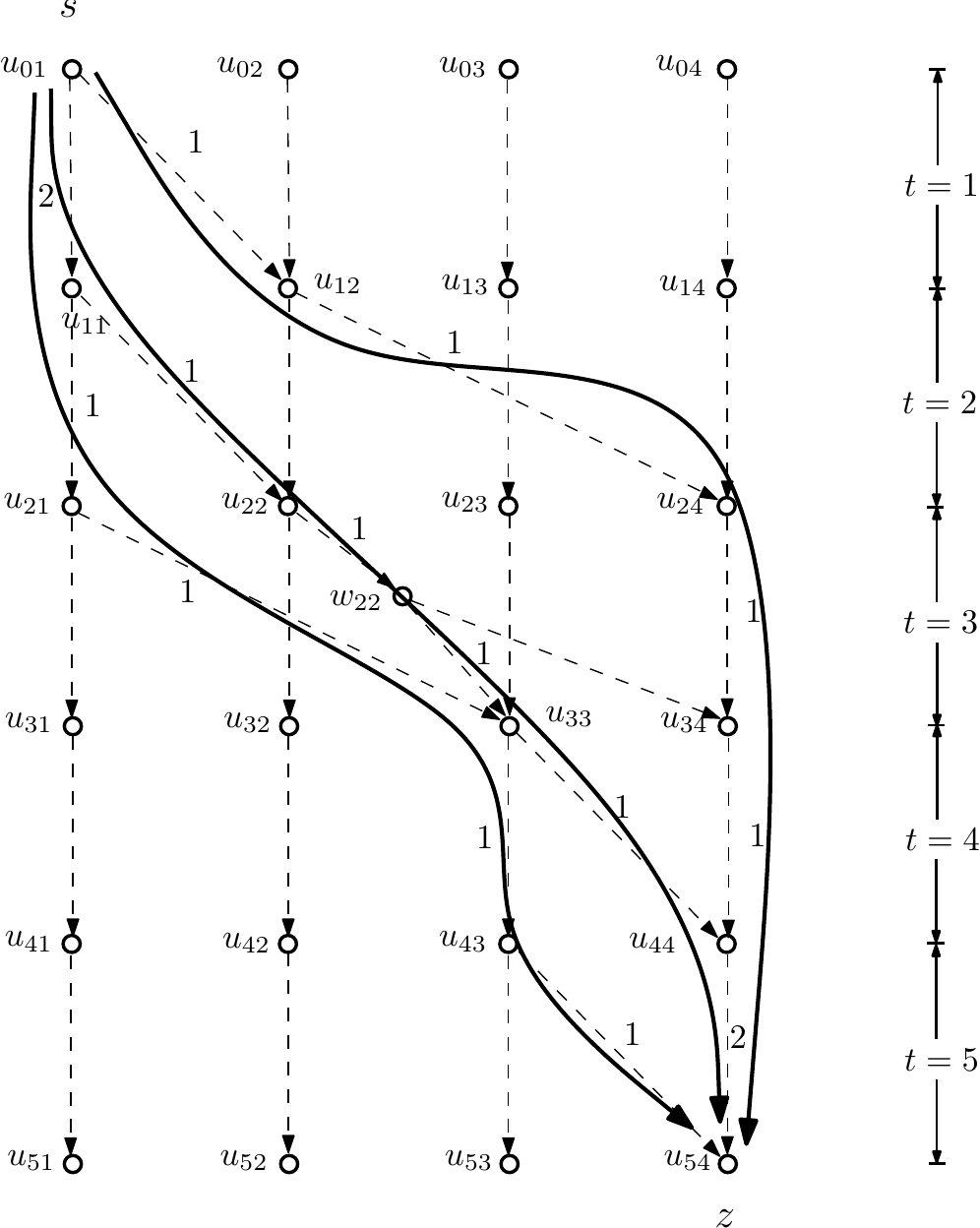}
        \label{fig:shape4}}        
        }
   \caption{(a) The static expansion of a temporal graph. Here, only two edges leave from the same node at the same time: $(u_{22},u_{33})$ and $(u_{22},u_{34})$. (b) Adding a new node $w_{22}$ and three new edges. This we ensures that a node departure time can be removed by removing a single diagonal edge: removing edge $(u_{22},w_{22})$ removes all possible departures from $u_{22}$. This ensures that separation of $s$ and $z$ by node-departure times is equivalent to separation by a usual static cut. (c) Adding capacities to the edges. Vertical edges take capacity $\lambda_{max}=5$ and diagonal edges take capacity 1. (d) The maximum number of out-disjoint journeys from $s$ to $z$ is equal to the maximum flow from $s$ to $z$ and both are equal to 3.} \label{fig:menger-example-expansion}
\end{figure}

The idea is to take the static expansion $H=(S,A)$ of $\lambda(G)$ and, for each time-node $u_{ij}$ with at least two outgoing edges to nodes different than $u_{i+1}j$, add a new node $w_{ij}$ and the edges $(u_{ij},w_{ij})$ and $(w_{ij},u_{(i+1)j_1}),$ $(w_{ij},u_{(i+1)j_2}),\ldots,(w_{ij},u_{(i+1)j_k})$. Then define an \emph{edge capacity function $c:A\rightarrow \{1,\lambda_{\max}\}$} as follows: edges $(u_{ij},u_{(i+1)j})$ take capacity $\lambda_{\max}$ and all other edges take capacity $1$. The theorem follows by observing that the maximum $u_{01}$-$u_{\lambda_{\max}n}$ flow is equal to the minimum of the capacity of a $u_{01}$-$u_{\lambda_{\max}n}$ cut, the maximum number of out-disjoint journeys from $s$ to $z$ is equal to the maximum $u_{01}$-$u_{\lambda_{\max}n}$ flow, and the minimum number of node departure times needed to separate $s$ from $z$ is equal to the minimum of the capacity of a $u_{01}$-$u_{\lambda_{\max}n}$ cut. See also Figure \ref{fig:menger-example-expansion} for an illustration.

\section{Dissemination and Gathering of Information}
\label{sec:dissemination}

A natural application domain of temporal graphs is that of \emph{gossiping} and in general of \emph{information dissemination}, mainly by a distributed set of entities (e.g. a group of people or a set of distributed processes). Two early such examples were the \emph{telephone problem} \cite{BS72} and the \emph{minimum broadcast time problem} \cite{Ra94}. In both, the goal is to transmit some information to every participant of the system, while minimizing some measure of communication or time. A more modern setting, but in the same spirit, comes from the very young area of distributed computing in highly dynamic networks \cite{OW05,KLO10,KO11,CFQS12,MCS14,MCS13b}. 

There are $n$ nodes. In this context, nodes represent distributed processes. Note, however, that most of the results that we will discuss, concern centralized algorithms (and in case of lower bounds, these immediately hold for distributed algorithms as well). The nodes communicate with other nodes in discrete rounds by interchanging messages. In every round, an adversary scheduler selects a set of edges between the nodes and every node may communicate with its current neighbors, as selected by the adversary, usually by \emph{broadcasting} a single message to be delivered to all its neighbors. So, the dynamic topology behaves as a discrete temporal graph where the $i$-th instance of the graph is the topology selected by the adversary in round $i$. The main difference, compared to the setting of the previous sections, is that now (in all results that we will discuss in this section, apart from the last one) the topology is revealed to the algorithms in an online and totally unpredictable way. An interesting special case of temporal graphs consists of those temporal graphs that have \emph{connected instances}. A temporal graph $D$ is called \emph{continuously connected} (also known as \emph{1-interval connected}) if $D(t)$ is connected for all times $t\geq 1$ \cite{OW05,KLO10}. Such temporal graphs have some very useful properties concerning information propagation in a distributed setting, like, for example, that if all nodes broadcast in every round all information that they have heard so far, then in every round at least one more node learns something new, which implies that a piece of information can in principle be disseminated in at most $n-1$ rounds. Naturally, the problem of information dissemination becomes much more interesting and challenging if we do not allow nodes to transmit an unlimited amount of information in every round, that is, if we restrict the size of the messages that they can transmit.

An interesting problem of token dissemination in such a setting, called the $k$-\emph{token dissemination problem}, was introduced and first studied in \cite{KLO10}. In this problem there is a domain of \emph{tokens} $\calt$, each node is assigned a subset of the tokens, and a total of $k$ distinct tokens is assigned to the nodes. The goal is for an algorithm (centralized or distributed) to organize the communication between the nodes in such a way that, under any dynamic topology (from those described above), each node eventually terminates and outputs (i.e. has learned) all $k$ tokens. In particular, the focus here is on \emph{token-forwarding algorithms}. Such an algorithm is quite restricted in that, in every round $r$ and for every node $u$, it only picks a single token from those already known by $u$ (or the empty token $\perp$) and this token will be delivered to all the current neighbors of $u$ by a single broadcast transmission. Token-forwarding algorithms are simple, easy to implement, typically incur low overhead, and have been extensively studied in static networks \cite{Le92,Pe00}. We will present now a lower bound from \cite{KLO10} on the number of rounds for token dissemination, that holds even for centralized token forwarding algorithms. Such centralized algorithms are allowed to see and remember the whole state and history of the entire network, but they have to make their selection of tokens to be forwarded without knowing what topology will be scheduled by the adversary in the current round. So, first the algorithm selects and then the adversary reveals the topology, taking into account the algorithm's selection. For simplicity, it may be assumed that each of the $k$ tokens is assigned initially to exactly one (distinct) node.  

\begin{theorem} [\cite{KLO10}] \label{the:klo10-lower-bound}
Any deterministic centralized algorithm for $k$-token dissemination in continuously connected temporal graphs requires at least $\Omega(n\log k)$ rounds to complete in the worst case.
\end{theorem}

The idea behind the proof is to define a potential function that charges by $1/(k-i)$ the $i$-th token learned by each node. So, for example, the first token learned by a node comes at a cheap price of $1/k$ while the last token learned costs $1$. The initial total potential is $1$, because $k$ nodes have obtained their first token each, and the final potential (i.e. when all nodes have learned all $k$ tokens) is $n\cdot H_k=\Theta(n\log k)$. Then it suffices to present an adversarial schedule, i.e. a continuously connected temporal graph, that forces any algorithm to achieve in every round at most a bounded increase in potential. The topology of a round can be summarized as follows. First we select all edges that contribute no cost, called \emph{free edges}. An edge $\{u,v\}$ is free if the token transmitted by $u$ is already known by $v$ and vice versa. The free edges partition the nodes into $l$ components $C_1,C_2,\ldots,C_l$. We pick a representative $v_i$ from each component $C_i$. It remains to construct a connected graph over the $v_i$s. An observation is that each $v_i$ transmits a distinct token $t_i$, otherwise at least two of them should have been connected by a free edge (because two nodes interchanging the same token cannot learn anything new). The idea is to further partition the representatives into a small set of nodes that know many tokens each and a large set of nodes that know few tokens each. We can call the nodes that know many tokens the \emph{expensive} ones, because according to the above potential function a new token at a node that already knows a lot of tokens comes at a high price, and similarly we call those nodes that know few tokens the \emph{cheap} ones. In particular, a node is expensive if it is missing at most $l/6$ tokens and cheap otherwise. Roughly, a cheap node learns a new token at the low cost of at most $6/l$, because the cost of a token is inversely proportional to the number of missing tokens before the token's arrival. First we connect the cheap nodes by an arbitrary line. As there are at most $l$ such nodes and each one of them obtains at most two new tokens (because it has at most two neighbors on the line and each node transmits a single token), the total cost of this component is at most $12$, that is, bounded as desired. It remains to connect the expensive nodes. It can be shown that there is a way to match each expensive node to a distinct cheap node (i.e. by constructing a matching between the expensive and the cheap nodes), so that no expensive node learns a new token. So, the only additional cost is that of the new tokens that cheap nodes obtain from expensive nodes. This additional cost is roughly at most $6$, so the total cost have been shown to be bounded by a small constant as required. It is worth mentioning that \cite{KLO10}, apart from the above lower bound, also proposed a simple distributed algorithm for $k$-token dissemination that needs $O(nk)$ rounds in the worst case to deliver all tokens.

The above lower bound can be further improved by exploiting the probabilistic method \cite{DPRS13}. In particular it can be shown that any randomized token-forwarding algorithm (centralized or distributed) for $k$-token dissemination needs $\Omega(nk/\log n)$ rounds. This lower bound is within a logarithmic factor of the $O(nk)$ upper bound of \cite{KLO10}. The construction is trivial (as is typical in probabilistic results, the interesting machinery is in the analysis). Now \emph{all} the representatives of the connected components formed by the free edges are connected arbitrarily by a line. The idea is to first prove the bound w.h.p. over an initial token distribution, in which each of the nodes receives each of the $k$ tokens independently with probability $3/4$. It can be shown in this case that, w.h.p. over the initial assignment of tokens, in every round there are at most $O(\log n)$ new token deliveries and an overall of $\Omega(nk)$ new token deliveries must occur for the protocol to complete. Finally, it can be shown via the probabilistic method, that, in fact, any initial token distribution can be reduced to the above distribution for which the bound holds. The above lower bounding technique, based on the probabilistic method, was applied in \cite{HK12} to several variations of $k$-token dissemination. For example, if the nodes are allowed to transmit $b\leq k$ tokens instead of only one token in every round, then it can be proved that any randomized token-forwarding algorithm requires $\Omega(n + nk/(b^2 \log n \log \log n))$ rounds.

In \cite{DPRS13}, also offline token forwarding algorithms were designed, that is, algorithms provided the whole dynamic topology in advance. One of the problems that they studied, was that of delivering all tokens to a given sink node $z$ as fast as possible, called the \emph{gathering problem}. We now present a lemma from \cite{DPRS13} concerning this problem, mainly because its proof constitutes a nice application of the temporal analogue of Menger's theorem presented in Section \ref{sec:menger} (the simplified proof via Menger's temporal analogue is from \cite{MMCS13}).

\begin{lemma} [DPRS13]
Let there be $k\leq n$ tokens at given source nodes and let $z$ be an arbitrary node. Then, if the temporal graph $D$ is continuously connected, all the tokens can be delivered to $z$, using local broadcasts, in $O(n)$ rounds.
\end{lemma}

Let $S=\{s_1,s_2,\ldots,s_h\}$ be the set of source nodes, let $N(s_i)$ be the number of tokens of source node $s_i$ and let the age of the temporal graph be $n+k=O(n)$. It suffices to prove that there are at least $k$ out-disjoint journeys from $S$ to any given $z$, such that $N(s_i)$ of these journeys leave from each source node $s_i$. Then, all tokens can be forwarded in parallel, each on one of these journeys, without conflicting with each other in an outgoing transmission and, as the age is $O(n)$, they all arrive at $z$ in $O(n)$ rounds. To show the existence of $k$ out-disjoint journeys, we create a \emph{supersource} node $s$ and connect it to the source node with token $i$ (assuming an arbitrary ordering of the tokens from $1$ to $k$) by an edge labeled $i$. Then we shift the rest of the temporal graph in time, by increasing all other edge labels by $k$. The new temporal graph $D^\prime$ has asymptotically the same age as the original and all properties have been preserved. Now, it suffices to show that there are at least $k$ out-disjoint journeys from $s$ to $z$, because the $k$ edges of $s$ respect the $N(s_i)$'s. Due to Menger's temporal analogue, it is equivalent to show that at least $k$ departure times must be removed to separate $s$ from $z$. Indeed, any removal of fewer than $k$ departure times must leave at least $n$ rounds during which all departure times are available (because, due to shifting by $k$, the age of $D^\prime$ is $n+2k$). Due to the fact that the original temporal graph is connected in every round, $n$ rounds guarantee the existence of a journey from $s$ to $z$. 

\section{Design Problems}
\label{sec:design}

So far, we have mainly presented problems in which a temporal graph is provided somehow (either in an offline or an online way) and the goal is to solve a problem on that graph. Another possibility is when one wants to \emph{design} a desired temporal graph. In most cases, such a temporal graph cannot be arbitrary, but it has to satisfy some properties prescribed by the underlying application. This design problem was introduced and studied in \cite{MMCS13} (and its full version \cite{MMS15}). An abstract definition of the problem is that we are given an underlying (di)graph $G$ and we are asked to assign labels to the edges of $G$ so that the resulting temporal graph $\lambda (G)$ minimizes some parameter while satisfying some connectivity property. The parameters studied in \cite{MMCS13} were the maximum number of labels of an edge, called the \emph{temporality}, and the total number of labels, called the \emph{temporal cost}. The connectivity properties of \cite{MMCS13} had to do with the preservation of a subset of the paths of $G$ in time-respecting versions. For example, we might want to preserve all reachabilities between nodes defined by $G$, in the sense that for every pair of nodes $u,v$ such that there is a path from $u$ to $v$ in $G$ there must be a temporal path from $u$ to $v$ in $\lambda (G)$. Another such property is to guarantee in $\lambda (G)$ time-respecting versions of all possible paths of $G$. All these can be thought of as trying to \emph{preserve} a connectivity property of a static graph in the temporal dimension while trying to minimize some cost measure of the resulting temporal graph. 

The provided graph $G$ represents some given static specifications, for example the available roads between a set of cities or the available routes of buses in the city center. In scheduling problems it is very common to have such a static specification and to want to organize a temporal schedule on it, for example to specify the precise time at which a bus should pass from a particular bus stop while guaranteeing that every possible pair of stops are connected by a route. Furthermore, it is very common that any such solution should at the same time take into account some notion of cost. Minimizing cost parameters may be crucial as, in most real networks, making a connection available and maintaining its availability does not come for free. For example, in wireless sensor networks the cost of making edges available is directly related to the power consumption of keeping nodes awake, of broadcasting, of listening the wireless channel, and of resolving the resulting communication collisions. The same holds for transportation networks where the goal is to achieve good connectivity properties with as few transportation units as possible.

\begin{figure}[!hbtp]
\centering{
\includegraphics[width=0.4\textwidth]{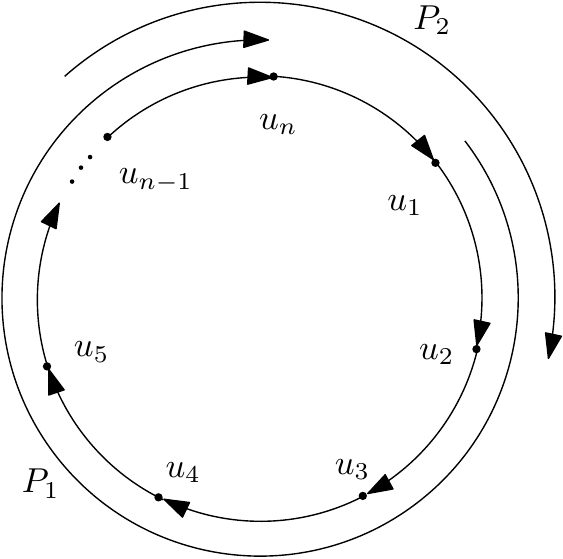}
}
\caption{Path $P_2$ forces a second label to appear on either $(u_{n-1},u_n)$ or $(u_1,u_2)$.} \label{fig:ring}
\end{figure}

For an example, imagine that we are given a directed ring $u_1,u_2,\ldots,u_n$ and we want to assign labels to its edges so that the resulting temporal graph has a journey for every simple path of the ring and at the same time minimizes the maximum number of labels of an edge. In more technical terms, we want to determine or bound the \emph{temporality} of the ring subject to the \emph{all paths} property. It is worth mentioning that the temporality (and the temporal cost) is defined as the \emph{minimum} possible achievable value that satisfies the property, as, for example, is also the case for the chromatic number of a graph, which is defined as the minimum number of colors that can properly color a graph. Looking at Figure~\ref{fig:ring}, it is immediate to observe that an increasing sequence of labels on the edges of path $P_1$ implies a decreasing pair of labels on edges $(u_{n-1},u_n)$ and $(u_1,u_2)$. On the other hand, path $P_2$ uses first $(u_{n-1},u_n)$ and then $(u_1,u_2)$ thus it requires an increasing pair of labels on these edges. It follows that in order to preserve both $P_1$ and $P_2$ we have to use a second label on at least one of these two edges, thus the temporality is at least 2. Next, consider the labeling that assigns to each edge $(u_i,u_{i+1})$ the labels $\{i, n+i\}$, where $1\leq i\leq n$ and $u_{n+1}=u_1$. It is not hard to see that this labeling preserves all simple paths of the ring. Since the maximum number of labels that it assigns to an edge is 2, we conclude that the temporality is also at most 2. Taking both bounds into account, we may conclude that the temporality of preserving all simple paths of a directed ring is 2. Moreover, it holds that the temporality of graph $G$ is lower bounded by the maximum temporality of its subgraphs, because if a labeling preserves all paths of $G$ then it has to preserve all paths of any subgraph of $G$, paying every time the temporality of the subgraph. So, for example, if the input graph $G$ contains a directed ring then the temporality of $G$ must be at least $2$ (and could be higher depending on the structure of the rest of the graph).   

\begin{figure}[!hbtp]
   \centering{
        \includegraphics[width=0.8\textwidth]{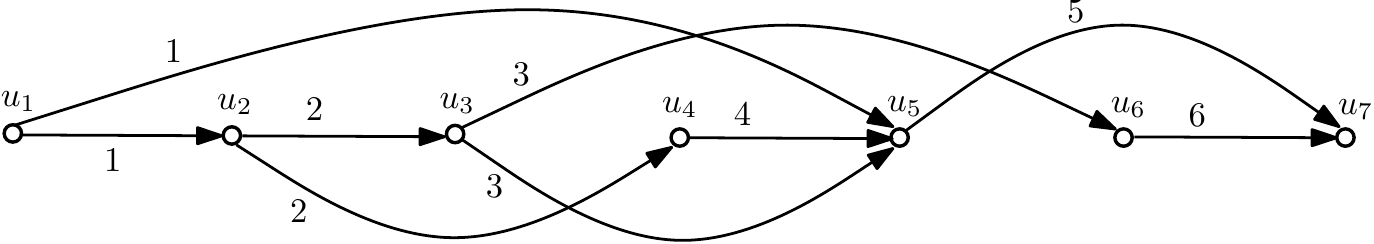}
        }
   \caption{A topological sort of a DAG. Edges are labeled by the indices of their tails (which are strictly increasing from left to right) and this labeling converts every possible path of the dag to a journey. For example, $(u_1,u_2,u_3,u_5,u_7)$ is a journey because its labels $(1,2,3,5)$ are strictly increasing.} \label{fig:dag}
\end{figure}

Rings have very small temporality w.r.t. the all paths property, however there is a large family of graphs with even smaller. This is the family of directed acyclic graphs (DAGs). DAGs have the very convenient property that they can be topologically sorted. In fact, DAGs are the only digraphs that satisfy this property. A \emph{topological sort} of a digraph $G$ is a linear ordering of its nodes such that if $G$ contains an edge $(u,v)$ then $u$ appears before $v$ in the ordering. So, we can order the nodes from left to right and have all edges pointing to the right.  Now, we can assign to the nodes the indices $1,2,\ldots,n$ in ascending order from left to right and then assign to each edge the label of its tail, as shown in Figure \ref{fig:dag}. In this way, every edge obtains exactly one label and every path of $G$ has been converted to a journey, because every path moves from left to right thus always moves to greater node indices. As these indices are also the labels of the corresponding edges, the path has strictly increasing labels which makes it a journey. This, together with the fact that the temporality is at least 1 in all graphs with non-empty edge sets, shows that the temporality of any DAG w.r.t. the all paths property is $1$.

\begin{figure}[!hbtp]
\centering{
\includegraphics[width=0.9\textwidth]{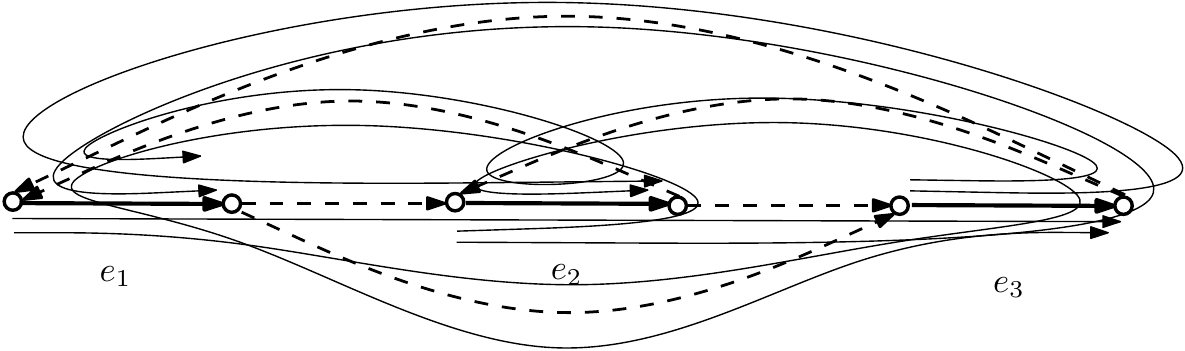}
}
\caption{The graph consists of the solid and dashed edges. The long curves highlight some of the paths that the graph defines. Edges $e_1$, $e_2$, and $e_3$ constitute an edge-kernel of the graph, because for every possible permutation of these edges the graph has a directed path (one of those highlighted in the figure) that traverses the edges in the order defined by the permutation. As a result, at least 3 labels must be assigned on an edge in order to preserve a temporal analogue of every possible path.} \label{fig:edge-kernel}
\end{figure}

In both of the above examples, all paths could be preserved by using very few labels per edge. One may immediately wonder whether converting all paths to journeys can always be achieved with few labels per edge, e.g. a constant number of labels. However, a more careful look at the previous examples may provide a first indication that this is not the case. In particular, the ring example suggests that cycles can cause an increase of temporality, compared to graphs without cycles, like DAGs. Of course, a single ring only provides a very elementary exposition of this phenomenon, however as proved in \cite{MMCS13}, this core observation can be extended to give a quite general method for lower bounding the temporality. The idea is to identify a subset of the edges of $G$ such that, for every possible permutation of these edges, $G$ has a path following the direction of the permutation. Such subsets of edges, with many interleaved cycles, are called \emph{edge-kernels} (see Figure \ref{fig:edge-kernel} for an example) and it can be proved that the preservation of all paths of an edge-kernel on $k$ edges yields a temporality of at least $k$. To see this, consider an edge-kernel $K=\{e_1,e_2,\ldots,e_k\}$ and order increasingly the labels of each edge. Now take an edge with maximum first label, move from it to an edge of maximum second label between the remaining edges, then move from this to an edge of maximum third label between the remaining edges, and so on. All these moves can be performed because $K$ is an edge-kernel, thus there is a path no matter which permutation of the edges we choose. As in step $i$ we are on the edge $e$ with maximum $i$-th label, we cannot use the 1st, 2nd, $\ldots$, $i$-th labels of the next edge to continue the journey because none of these can be greater than the $i$-th label of $e$. So, we must necessarily use the $(i+1)$-th label of the next edge, which by induction shows that in order to go through the $k$-th edge in this particular permutation we need to use a $k$th label on that edge.

Also, as stated above, the temporality of a graph w.r.t to the all paths property is always lower bounded by the temporality of any of its subgraphs. As a consequence, we can obtain a lower bound on the temporality of a graph or of a whole graph family by identifying a large edge-kernel in it. For a simple application of this method, it is possible to show that in order to preserve all paths of a complete digraph, at least $\lfloor n/2\rfloor$ labels are required on some edge. This is done by showing that complete digraphs have an edge-kernel of size $\lfloor n/2\rfloor$. Moreover, it is possible to construct a planar graph containing an edge-kernel of size $\Omega(n^{1/3})$, which yields that there exist planar graphs with temporality at least $\Omega(n^{1/3})$. It is worth noting that the absence of a large edge-kernel does not necessarily imply small temporality. In fact, it is an interesting open problem whether there are other structural properties of the underlying graph that could cause a growth of the temporality.

The above show that preserving all paths in time can be very costly in several cases. On the other hand, preserving  only the reachabilities can always be achieved inexpensively. In particular, it can be proved that for every strongly connected digraph $G$, we can preserve a journey from $u$ to $v$ for every $u,v$ for which there exists a path from $u$ to $v$ in $G$, by using at most two labels per edge \cite{MMCS13}. Recall the crucial difference: now it suffices to preserve \emph{a single path} from all possible paths that go from $u$ to $v$. The result is proved by picking any node $u$ and considering an in-tree rooted at $u$. We then label the edges of each level $i$, counting from the leaves, with label $i$, so that all paths of the tree become time-respecting (this also follows from the fact that the tree is a DAG so, as we discussed previously, all of its paths can be preserved with a single label per edge). Next we consider an out-tree rooted at $u$ and we label that tree inversely, i.e. from the root to the leaves, and beginning with the label $i+1$. The first tree has a journey from every node to $u$ arriving by time $i$ and the second tree has a journey from $u$ to every other node beginning at time $i+1$. This shows that there is a journey from every node to every other node. Moreover, this was achieved by using at most two labels per edge because every edge of the in-tree has a single label and every edge of the out-tree has a single label and an edge is in the worst case used by both trees, in which case it is assigned two labels. Furthermore, it can be proved that the temporality w.r.t. reachabilities of any digraph $G$ is upper bounded by the maximum temporality of its strongly connected components. But we just saw that each component needs at most two labels, thus it follows that two labels per edge are sufficient for preserving all reachabilities of \emph{any} digraph $G$. 

\begin{figure}[!hbtp]
\centering{
\includegraphics[width=0.8\textwidth]{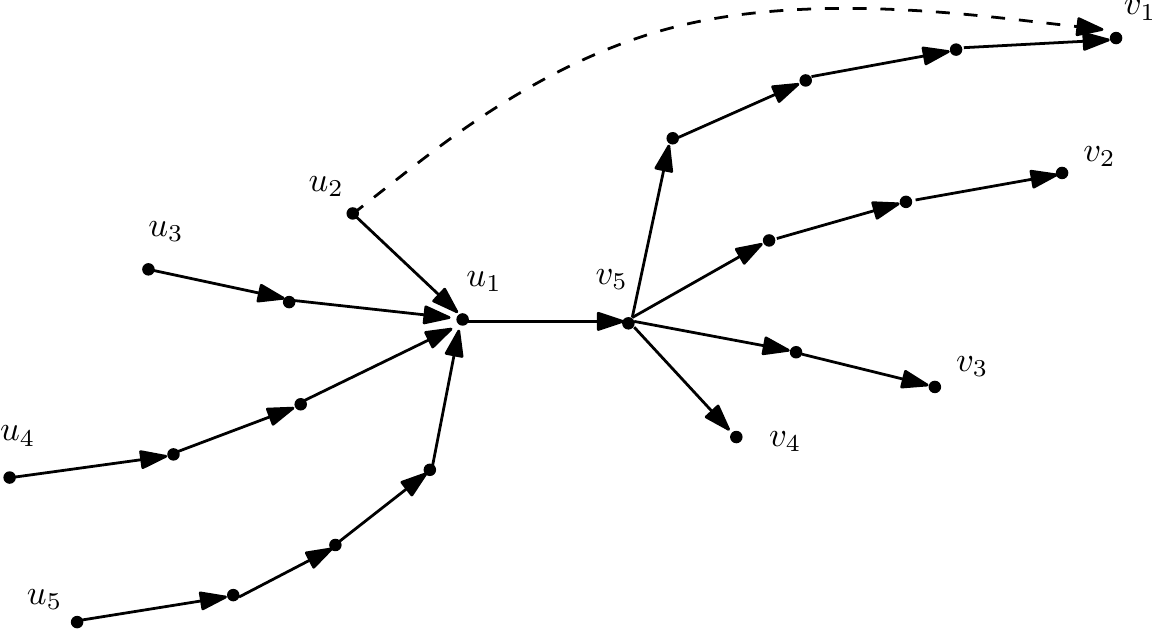}
}
\caption{In this example, restricting the maximum label to be at most equal to the diameter $d(G)$ forces the temporality to be at least $d(G)$.} \label{fig:diam}
\end{figure}

Finally, we should mention an interesting relation between the temporality and the age of a temporal graph. In particular, restricting the maximum label that the labeling is allowed to use makes the temporality grow. For an intuition why this happens, consider the case in which there are many maximum length shortest paths between different pairs of nodes that all must be necessarily be preserved in order to preserve the reachabilities. Now if it happens that all of them pass through the same edge $e$ but use $e$ at many different times, then $e$ must necessarily have many different labels, one for each of these paths. A simple example to further appreciate this is given in Figure \ref{fig:diam}. In that figure, each $u_i$-$v_i$ path is a unique shortest path between $u_i$ and $v_i$ and has additionally length equal to the diameter (i.e. it is also a maximum one), so we must necessarily preserve all 5 $u_i$-$v_i$ paths. Note now that each $u_i$-$v_i$ path passes through $e$ via its $i$-th edge. Each of these paths can only be preserved without violating $d(G)$ by assigning the labels $1,2,\ldots, d(G)$, however note that then edge $e$ must necessarily have all labels $1,2,\ldots,d(G)$. To see this, notice simply that if any label $i$ is missing from $e$ then there is some maximum shortest path that goes through $e$ at step $i$. As $i$ is missing it cannot arrive sooner than time $d(G)+1$ which violates the preservation of the diameter. Additionally, the following trade-off for the particular case of a ring can be proved \cite{MMCS13}: If $G$ is a directed ring and the age is $(n-1)+k$, then the temporality of preserving all paths is $\Theta(n/k)$, when $1\leq k \leq n-1$, and $n-1$, when $k=0$. 

\section{Temporal Versions of Other Standard Graph Problems: Complexity and Solutions}
\label{sec:matching}

Though it is not yet clear how is the complexity of combinatorial optimization problems affected by introducing to them a notion of time, still there is evidence that complexity increases significantly and that totally novel solutions have to be developed in several cases. In an early but serious attempt to answer the above question, Orlin \cite{Or81} observed that many dynamic languages derived from $\rem{NP}$-complete languages can be shown to be $\rem{PSPACE}$-complete. This increase in complexity has been also reported in \cite{BF03,XFJ03}. For example, \cite{BF03} studied the computation of multicast trees minimizing the overall transmission time and to this end proved that it is $\rem{NP}$-complete to compute strongly connected components in temporal graphs. Important evidence to this direction comes also from the rich literature on labeled graphs, a more general model than temporal graphs, with different motivation, and usually interested in different problems than those resulting when the labels are explicitly regarded as time moments. Several papers in this direction have considered labeled versions of polynomial-time solvable problems, in which the goal is to minimize/maximize the number of labels used by a solution. For example, the first labeled problem introduced in the literature was the {\sc Labeled Minimum Spanning Tree} problem, which has several applications in communication network design. This problem is $\rem{NP}$-hard and many complexity and approximability results have been proposed (see e.g. \cite{BL97,KW98}). On the other hand, the {\sc Labeled Maximum Spanning Tree} problem has been shown polynomial in \cite{BL97}. In \cite{BLWZ05}, the authors proved that the {\sc Labeled Minimum Path} problem is $\rem{NP}$-hard and provided some exact and approximation algorithms. In \cite{Mo05}, it was proved that the {\sc Labeled Perfect Matching} problem in bipartite graphs is $\rem{APX}$-complete (see also \cite{TIR78} for a related problem). 

A primary example of this phenomenon, of significant increase in complexity when extending a combinatorial optimization problem in time, is the fundamental {\sc Maximum Matching} problem. In its static version, we are given a graph $G=(V,E)$ and we must compute a maximum cardinality set of edges such that no two of them share an endpoint. {\sc Maximum Matching} can be solved in polynomial time by the famous Edmonds' algorithm \cite{Ed65} (the time is $O(\sqrt{|V|}\cdot |E|)$ by the algorithm of \cite{MV80}). Now consider the following temporal version of the problem, called {\sc Temporal Matching} in \cite{MS14}. In this problem, we are given a temporal graph $D=(V,A)$ and we are asked to decide whether there is a maximum matching $M$ of the underlying static graph of $D$ that can be made temporal by selecting a single label $l\in\lambda(e)$ for every edge $e\in M$. For a single-labeled matching to be temporal it suffices to guarantee that no two of its edges have the same label. {\sc Temporal Matching} was proved in \cite{MS14} to be $\rem{NP}$-complete. Then the problem of computing a maximum cardinality temporal matching is immediately $\rem{NP}$-hard, because if we could compute such a maximum temporal matching in polynomial time, we could then compare its cardinality to the cardinality of a maximum static matching and decide {\sc Temporal Matching} in polynomial time. $\rem{NP}$-completeness of {\sc Temporal Matching} can be proved by the sequence of polynomial-time reductions: {\sc Balanced 3SAT} $\leq_P$ {\sc Balanced Union Labeled Matching} $\leq_P$ {\sc Temporal Matching}. In {\sc Balanced 3SAT}, which is known to be $\rem{NP}$-complete, every variable $x_i$ appears $n_i$ times negated and $n_i$ times non-negated and in {\sc Balanced Union Labeled Matching} we are given a bipartite graph $G=((X,Y),E)$, labels $L=\{1,2,...,h\}$, and a labeling $\lambda : E \rightarrow 2^L$, every node $u_i\in X$ has precisely two neighbors $v_{ij}\in Y$, and additionally both edges of $u_i$ have the same number of labels, and we must decide whether there is a maximum matching $M$ of $G$ s.t. $\bigcup_{e\in M} \lambda(e) = L$ \cite{MS14}.

Another interesting problem is the {\sc Temporal Exploration} problem \cite{MS14}. In this problem, we are given a temporal graph and the goal is to visit all nodes of the temporal graph by a temporal walk, that possibly revisits nodes, minimizing the arrival time. The version of this problem for static graphs is well-known as {\sc Graphic TSP}. Though, in the static case, the decision version of the problem, asking whether a given graph is explorable, can be solved in linear time, in the temporal case it becomes $\rem{NP}$-complete. Additionally, in the static case, there is a $(3/2-\varepsilon)$-approximation for undirected graphs \cite{GSS11} and a $O(\log n/\log\log n)$ for directed \cite{AGMGS10}. 

\begin{figure}[!hbtp]
   \centering{
        \subfigure[]{
        \includegraphics[width=0.8\textwidth]{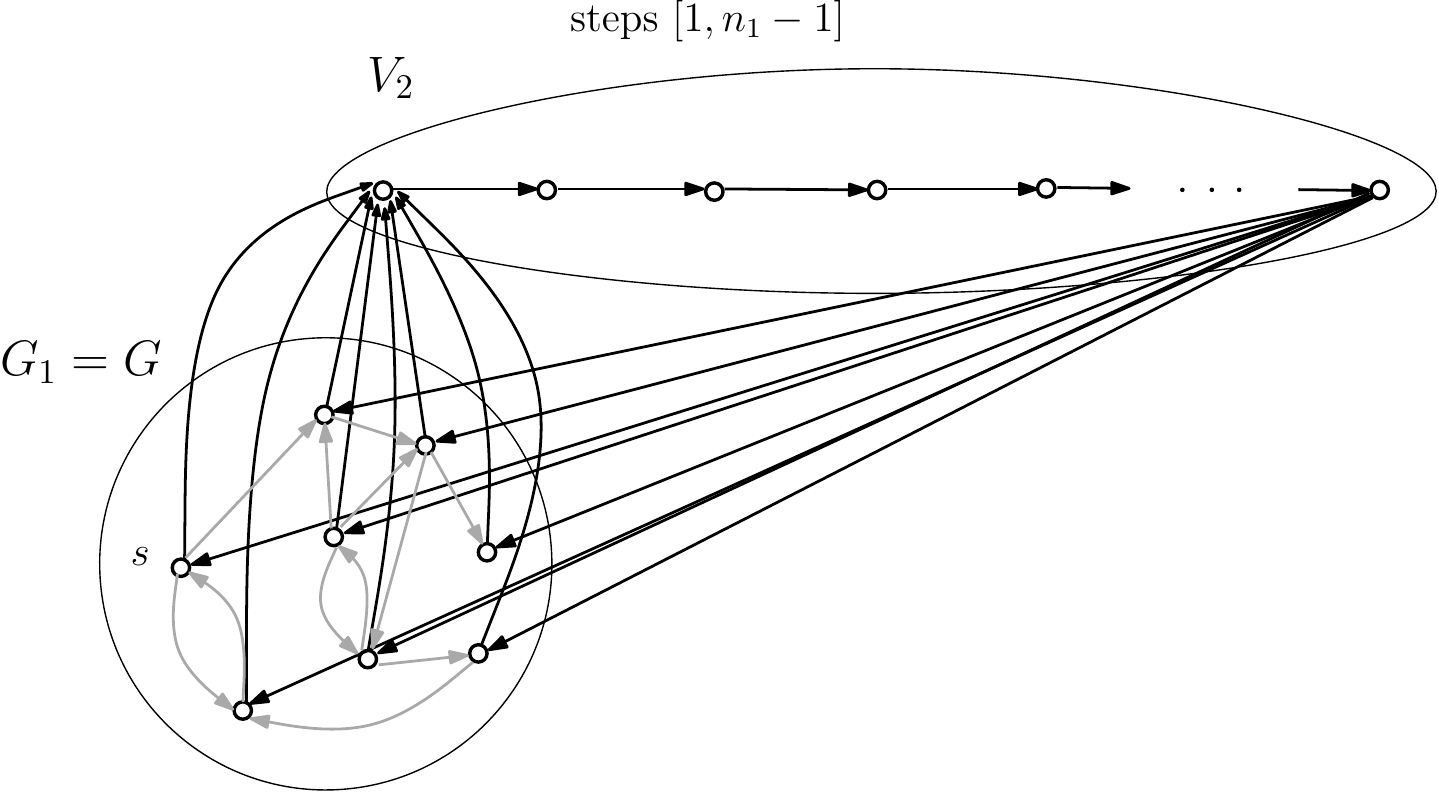}
        \label{fig:oneint1}}
	\hspace{1cm}
        \subfigure[]{
        \includegraphics[width=0.8\textwidth]{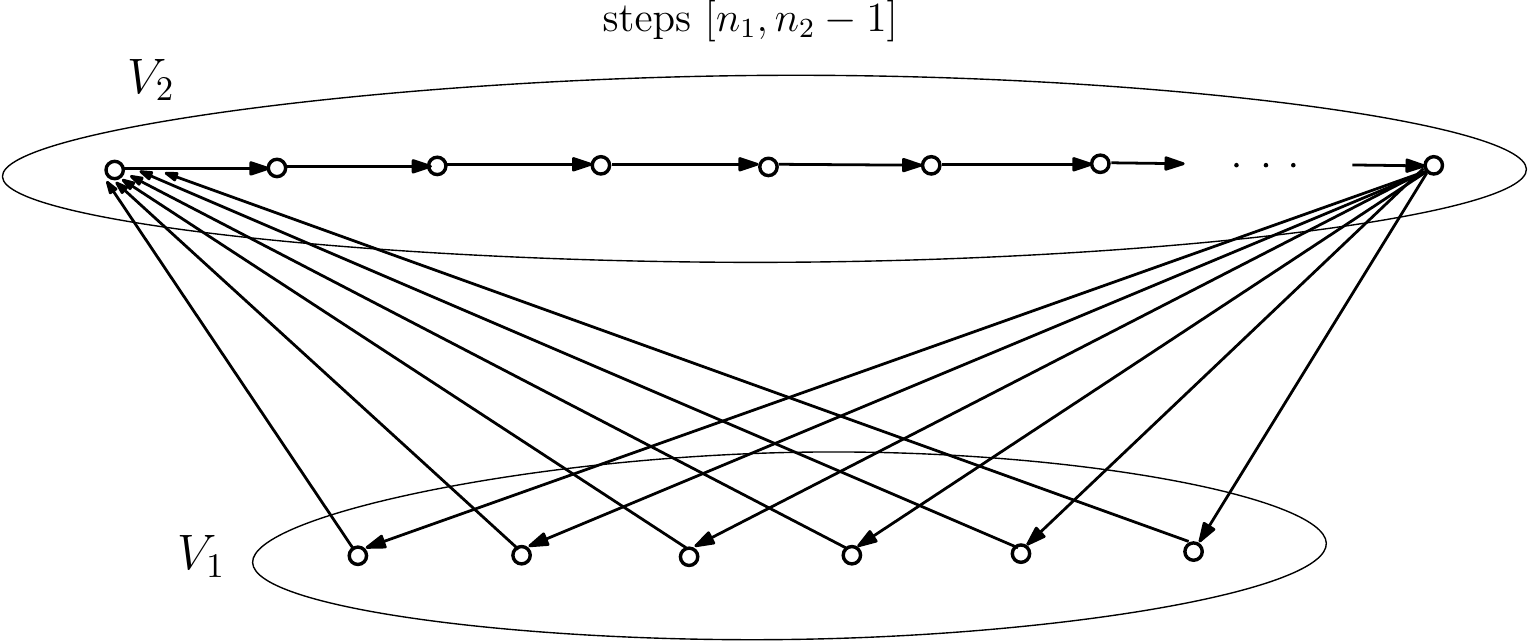}
        \label{fig:oneint2}}
	\hspace{1cm}
        \subfigure[]{
        \includegraphics[width=0.8\textwidth]{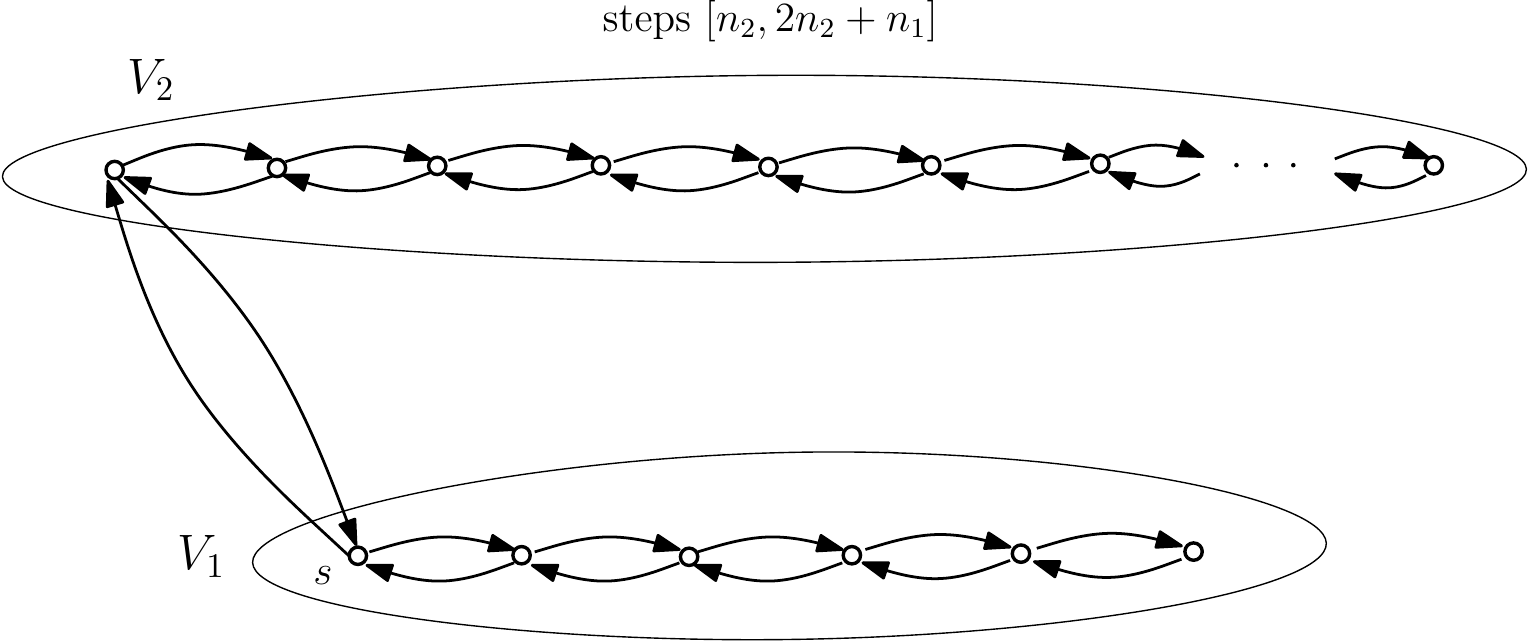}
        \label{fig:oneint3}}
        }
   \caption{The temporal graph constructed by the reduction. (a) $T_1$ (b) $T_2$ (c) $T_3$} \label{fig:oneint}
\end{figure}

In contrast to these, it was proved in \cite{MS14} that there exists some constant $c>0$ such that {\sc Temporal Exploration} cannot be approximated within $cn$ unless $\rem{P}=\rem{NP}$, by presenting a gap introducing reduction from {\sc Hampath}. Additionally, it was proved that even the special case in which every instance of the temporal graph is connected, cannot be approximated within $(2-\varepsilon)$, for every constant $\varepsilon>0$, unless $\rem{P}=\rem{NP}$. The reduction is from {\sc Hampath} (input graph $G$, source $s$). The constructed temporal graph $D$ consists of three strongly connected static graphs $T_1$, $T_2$, and $T_3$ persisting for the intervals $[1,n_1-1]$, $[n_1,n_2-1]$, and $[n_2,2n_2+n_1]$, respectively (it will be helpful at this point to look at Figure \ref{fig:oneint}). We can restrict attention to instances of {\sc Hampath} of order at least $2/\varepsilon$, without affecting its $\rem{NP}$-completeness. We also set $n_2=n_1^2+n_1$ (in fact, we can set $n_2$ equal to any polynomial-time computable function of $n_1$). If $G$ is hamiltonian, then for the arrival time, $\OPT$, of an optimum exploration it holds that $\OPT=n_1+n_2-1=n_1^2+2n_1-1$ while if $G$ is not hamiltonian, then $\OPT\geq$  $2n_2+1=$ $2(n_1^2+n_1)+1>$ $2(n_1^2+n_1)$, which can be shown to introduce the desired $(2-\varepsilon)$ gap. This negative result has been recently improved by Erlebach \emph{et al.} \cite{EHK15} to $O(n^{1-\varepsilon})$ for any $\varepsilon>0$. In the same work, an explicit construction of continuously connected temporal graphs that require $\Theta(n^2)$ steps to be explored was also given. 

On the positive side, it is not hard to show that in continuously connected temporal graphs, {\sc Temporal Exploration} can be approximated within the temporal diameter of the temporal graph \cite{MS14}. In \cite{EHK15}, the authors additionally studied the {\sc Temporal Exploration} problem in other interesting restricted families of temporal graphs, like temporal graphs in which the underlying graph has treewidth $k$ (a work explicitly concerned with the treewidth of temporal graphs and its relation to the treewidth of static graphs is \cite{MM13}), is a $2\times n$ grid, a cycle, a cycle with a chord, or a bounded-degree planar graph, for which they provided upper bounds on exploration time. See also \cite{FMS09} for another study of the exploration problem in temporal graphs with periodic edge-availabilities, from a distributed computing perspective.

\begin{figure}[!hbtp]
\centering{
\includegraphics[width=0.75\textwidth]{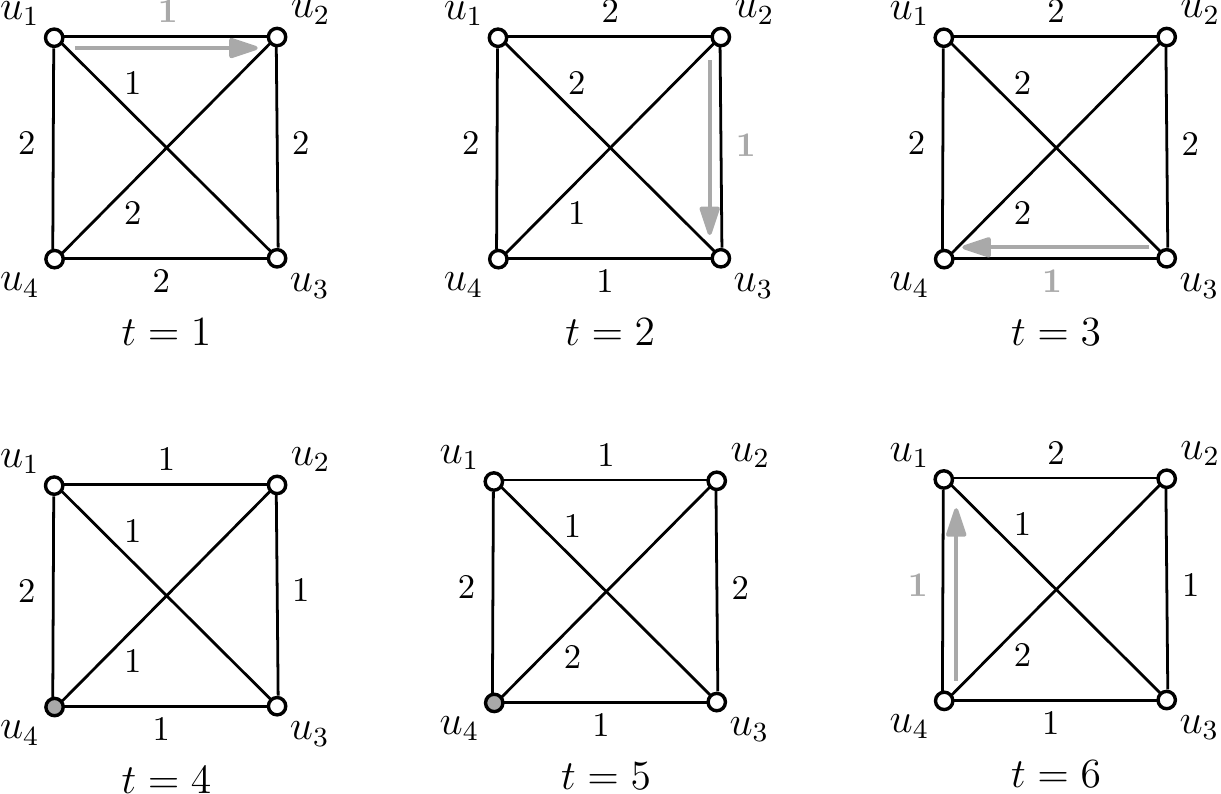}
}
\caption{An instance of TTSP(1,2) consisting of a complete temporal graph $D=(V,A)$, where $V=\{u_1,u_2,u_3,u_4\}$, and a cost function $c:A\rightarrow \{1,2\}$ which is presented by the corresponding costs on the edges. For simplicity, $D$ is an undirected temporal graph. Observe that the cost of an edge may change many times, e.g. the cost of $u_2u_3$ changes 5 times while of $u_1u_4$ changes only once. Here, the lifetime of the temporal graph is 6 and it is greater than $|V|$. The gray arcs and the nodes filled gray (meaning that the tour does not make a move and remains on the same node for that step) represent the TTSP tour $(u_1,1,u_2,2,u_3,3,u_4,6,u_1)$ that has cost $4=|V|$ and therefore it is an optimum TTSP tour.} \label{fig:tsp-example}
\end{figure}

Another demanding problem that becomes even more challenging in its temporal version is the famous {\sc Traveling Salesman Problem}, in which a graph with non-negative costs on its edge is provided and the goal is to find a tour visiting every node exactly once (called a \emph{TSP tour}), of minimum total cost. In one version of the problem, introduced in \cite{MS14}, the digraph remains static and complete throughout its lifetime but now each edge is assigned a cost that may change from instance to instance. So, the dynamicity has now been transferred from the topology to the costs of the edges. The goal is to find (by an offline centralized algorithm) a \emph{temporal TSP tour} of minimum total cost, where the cost of a tour is the sum of the costs of the time-edges that it traverses. The authors of \cite{MS14} introduced and studied the special case of this problem in which the costs are chosen from the set $\{1,2\}$. In particular, there is a cost function $c:A\rightarrow \{1,2\}$ assigning a cost to every time-edge of the temporal graph (see Figure \ref{fig:tsp-example} for an illustration). This is called the {\sc Temporal Traveling Salesman Problem with Costs One and Two} and abbreviated TTSP(1,2). Now observe that the famous (static) ATSP(1,2) problem is a special case of TTSP(1,2) when the lifetime of the temporal graph $D=(V,A)$ is restricted to $n$ and $c(e,t)=c(e,t^\prime)$ for all edges $e$ and times $t,t^\prime$. This immediately implies that TTSP(1,2) is also $\rem{APX}$-hard \cite{PY93} and cannot be approximated within any factor less than $207/206$ \cite{KS13} and the same holds for the interesting special case of TTSP(1,2) with lifetime restricted to $n$, that we will also discuss.

In the static case, one easily obtains a $(3/2)$-factor approximation for ATSP(1,2) by computing a perfect matching maximizing the number of ones and then patching the edges together arbitrarily. This works well, because such a minimum cost perfect matching can be computed in polynomial time in the static case by Edmonds' algorithm \cite{Ed65} and its cost is at most half the cost of an optimum TSP tour, as the latter consists of two perfect matchings. The $3/2$ factor follows because the remaining $n/2$ edges that are added during the patching process cost at most $n$, which, in turn, is another lower bound to the cost of the optimum TSP tour. This was one of the first algorithms known for ATSP(1,2). Other approaches have improved the factor to the best currently known $5/4$ \cite{Bl04}. Unfortunately, as we already discussed in the beginning of this section, even the apparently simple task of computing a matching maximizing the number of ones is not that easy in temporal graphs. A simple modification of those arguments yields that the problem remains $\rem{NP}$-hard if we require consecutive labels (in an increasing ordering) of the matching to have a time difference of at least two. Such time-gaps are necessary for constructing a time-respecting patching of the edges of the matching. In particular, if two consecutive edges of the matching had a smaller time difference, then the patching-edge would share time with at least one of them and the resulting tour would not have strictly increasing labels.

Our inability to compute a temporal matching in polynomial time, still does not exclude the possibility to find good approximations for it and then hope to be able to use them for obtaining good approximations for TTSP(1,2). Two main approaches were followed in \cite{MS14}. One was to reduce the problem to {\sc Maximum Independent Set} ({\sc MIS}) in $(k+1)$-claw free graphs and the other was to reduce it to {\sc $k^\prime$-Set Packing}, for some $k$ and $k^\prime$ to be determined. The first approach gives a $(7/4+\varepsilon)$-approximation ($=1.75+\varepsilon$) for the generic TTSP(1,2) and a $(12/7+\varepsilon)$-approximation ($\approx 1.71+\varepsilon$) for the special case of TTSP(1,2) in which the lifetime is restricted to $n$ (the latter is obtained by approximating a temporal path packing instead of a matching). The second approach improves these to $1.7+\varepsilon$ for the general case and to $13/8+\varepsilon=1.625+\varepsilon$ when the lifetime is $n$. In all the above cases, $\varepsilon>0$ is a small constant (not necessarily the same in all cases) adopted from the factors of the approximation algorithms for independent set and set packing.

We summarize now how the first of these approximations works. Consider the static expansion $H=(S,E)$ of $D$ and an edge $e=(u_{(i-1)j},u_{ij^\prime})\in E$. There are three types of conflicts, each defining a set of edges that cannot be taken together with $e$ in a temporal matching (with only unit time differences): (i) Edges of the same row as $e$, because these violate the unit time difference constraint (ii) edges of the same column as $u_{(i-1)j}$, because these share a node with $e$, thus violate the condition of constructing a matching, and (iii) edges of the same column as $u_{ij^\prime}$, for the same reason as (ii). Next consider the graph of edge conflicts $G=(E,K)$, where $(e_1,e_2)\in K$ iff $e_1$ and $e_2$ satisfy some of the above constraints (observe that the node set of $G$ is equal to the edge set of the static expansion $H$). Observe that temporal matchings of $D$ are now equivalent to independent sets of $G$. Moreover, $G$ is $4$-claw free meaning that there is no 4-independent set in the neighborhood of any node. To see that it is $4$-claw free, take any $e\in E$ and any set $\{e_1,e_2,e_3,e_4\}$ of four neighbors of $e$ in $G$. There are only 3 constraints thus at least two of the neighbors, say $e_i$ and $e_j$, must be connected to $e$ by the same constraint. But then $e_i$ and $e_j$ must also satisfy the same constraint with each other thus they are also connected by an edge in $G$. Now, from \cite{Ha95}, there is a factor of $3/5$ for {\sc MIS} in $4$-claw free graphs, which implies a $(3/5)$-approximation algorithm for temporal matchings. Simple modifications of the above arguments yield a $\frac{1}{2+\varepsilon}$-approximation algorithm for temporal matchings with time-differences at least two. Additionally, it can be proved that a $(1/c)$-factor approximation for the latter problem implies a $(2-\frac{1}{2c})$-factor approximation for TTSP(1,2). All these together, yield a $(7/4+\varepsilon)$-approximation algorithm for TTSP$(1,2)$ \cite{MS14}. 

An immediate question, which is currently open, is whether there is a $(3/2)$-factor approximation algorithm either for the general TTSP(1,2) or for its special case with lifetime restricted to $n$ (the reader may have observed that in the temporal case we have not yet achieved even the simplest factor of the static case).

\section{Linear Availabilities}
\label{sec:linear}

An interesting family of temporal graphs consists of those temporal graphs whose availability times are provided by some succinct representation. This could for example be a function, which we discuss here, or a probability distribution, which we discuss in the next section.

Such an example of a temporal graph in which a set of functions describes the availability times of the edges is the following (for other studies on periodically varying temporal graphs the reader is encouraged to consult \cite{Or81,FMS09,CFQS12} and references therein). The underlying graph is a complete static graph $G=(V,E)$. Each $e\in E$ has an associated linear function of the form $f_e(x)=a_ex+b_e$, where $x,a_e,b_e\in\bbbn_{\geq 0}$. For example, if an edge $e$ has $f_e(x)=3x+4$, then it is available at times $4,7,10,13,16,\ldots$. Clearly, the temporal graph that we obtain in this manner is $D=(V,A)$ where $A(r)=\{e\in E:f_e(x)=r$ for some $x\in\bbbn\}$. If we are additionally provided with a lifetime $l$ of the temporal graph then we just restrict $E(r)$ to $r\leq l$.

The above provides an immediate way for obtaining the $r$th instance for any $r$. For every $e\in E$, the $r$th instance contains edge $e$ iff $(r-b_e)/a_e$ is integer. It is important to note that, in the above family of temporal graphs, algorithmic solutions that depend at least linearly on the lifetime $l$ are not acceptable. The reason is that the lifetime $l$ is provided in binary so a linear dependence on $l$ grows exponentially in the binary representation of $l$. Foremost journeys in such graphs can be easily computed by a variation of the algorithm discussed in Section \ref{subsec:journeys}.

Now consider the following problem. We are given two edges $e_1$ and $e_2$ with corresponding functions $f_{e_1}(x)=a_1x+b_1$ and $f_{e_2}(x)=a_2x+b_2$ and we are asked to determine whether there is some instance having both edges, that is, to determine whether there exist $x_1$ and $x_2$ s.t. $f_{e_1}(x_1)=f_{e_2}(x_2)\lra$ $a_1x_1+b_1=a_2x_2+b_2\lra$ $a_1x_1=a_2x_2+(b_2-b_1)$. So, in fact, we are seeking for a $x_2$ s.t. $a_1\mid a_2x_2+(b_2-b_1)$ (where `$\mid$' reads as ``divides'') and we have reduced our problem to the problem of determining whether $c\mid ax+b$ for some $x$. Now imagine a right oriented ring of $c$ nodes numbered $0,1,\ldots,c-1$. Consider a process beginning from node $b \pmod c$ and making clockwise jumps of length $a$ in each round (where a round corresponds to an increment of $x$ by 1). We have that the process falls at some point on node $0$ iff $c\mid ax+b$ for some $x$. Viewed in this way, our problem is equivalent to checking whether $ax+b\equiv 0 \pmod c$ is solvable for the unknown $x$. This, in turn, may easily take the form $ax\equiv b^\prime \pmod c$ (given that $-b\equiv b^\prime \pmod c$) for $a>0$ and $c>0$ (equalities to 0 correspond to trivial cases of our original problem). Clearly, we have reduced our problem to the problem of detecting whether a modular linear equation admits a solution which is well-known to be solvable in polynomial time. In particular, a modular linear equation $ax\equiv b^\prime \pmod c$ has a solution iff $\gcd(a,c)\mid b^\prime$ (see e.g. \cite{CLRS01}, Corollary 31.21, page 869). Additionally, by solving the equation we can find all solutions modulo $c$ in $O(\log c + \gcd(a,c))$ arithmetic operations (see e.g. \cite{CLRS01}, page 871).

Note that in the case where $b_1=b_2=0$ then the answer to the problem is always ``yes'' as $a_1x_1=a_2x_2$ trivially holds for $x_1=a_2$ and $x_2=a_1$ (provided that $a_1a_2$ does not exceed the lifetime of the network if a lifetime is specified). In particular, if we are asked to determine the foremost instance containing both edges then this reduces to the computation of $\lcm(a_1,a_2)$ (where $\lcm$ is the least common multiple) which in turn reduces to the computation of $\gcd(a_1,a_2)$ by the equation $\lcm(a_1,a_2)=|a_1a_2|/\gcd(a_1,a_2)$. 

Now let us slightly simplify our model in order to obtain a solution to a more generic version of the above problem. We restrict the edge functions $a_ix+b_i$ so that $b_i<a_i$, e.g. $7x+4$. Then clearly, each such function corresponds to the whole equivalence subclass of $\bbbn$ modulo $a_i$ containing $b_i$, that is, $[b_i]_{a_i}=\{b_i+xa_i:x\in\bbbn\}$. So, for example, $7x+4$ corresponds to $\{4,11,18,25,\ldots\}$ in contrast to $7x+11$ that was allowed before and would just give the subset $\{11,18,25,\ldots\}$ of the actual class. Consider now the following problem: ``We are given a subset $E^\prime$ of the edge set $E$ and we want to determine whether there is some instance of the temporal graph containing all edges in $E^\prime$''. For simplicity, number the edges in $E^\prime$ from 1 to $k$. Formally, we want to determine the existence of some time $t$ s.t., for all $i\in \{1,2,\ldots,k\}$, there exists $x_i$ s.t. $t=a_ix_i+b_i$, or equivalently, $t\equiv b_i\pmod {a_i}$. Clearly, we have arrived at a set of simultaneous linear congruences and we can now apply the following known results.

\begin{theorem} [see e.g. \cite{BS96}, Theorem 5.5.5, pg 106] \label{the:bs1}
The system of congruences $t\equiv b_i\pmod {a_i}$, $1\leq i\leq k$, has a solution iff $b_i\equiv b_j\pmod {\gcd(a_i,a_j)}$ for all $i\neq j$. If the solution exists, it is unique modulo $\lcm(a_1,a_2,\ldots,a_k)$.
\end{theorem}

\begin{corollary} [see e.g. \cite{BS96}, Corollary 5.5.6, pg 106]
Let $a_1,a_2,\ldots,a_k$ be integers, each $\geq 2$, and define $a=a_1a_2\cdots a_k$, and $a^\prime = \lcm(a_1,a_2,\ldots,a_k)$. Given the system $S$ of congruences $t\equiv b_i\pmod {a_i}$, $1\leq i\leq k$, we can determine if $S$ has a solution, using $O(\lg^2 a)$ bit operations, and if so, we can find the unique solution modulo $a^\prime$, using $O(\lg^2 a)$ bit operations.
\end{corollary}

We may now return to the original formulation of our model in which $a_ix+b_i$ does not necessarily satisfy $b_i<a_i$. First keep in mind that $t_{min}=\max_{i\in E^\prime}\{b_i\}$ is the minimum time for every edge from $E^\prime$ to appear at least once (in fact, at that time, the last edge of $E^\prime$ appears). So we cannot hope to have them all in one instance sooner than this. Now notice that $a_ix+b_i$ is equivalent to $a_ix^\prime+(b_i\bmod a_i)$ for $x^\prime\geq \lfloor b_i/a_i\rfloor$; for example, $7x+15$ is equivalent to $7x^\prime+1$ for $x^\prime\geq 2$. In this manner, we obtain an equivalent setting in which again $b_i<a_i$ for all $i$ but additionally for every $i$ we have a constraint on $x$ of the form $x\geq q_i$. We may now ignore the constraints and apply Theorem \ref{the:bs1} to determine whether there is a solution to the new set of congruences as there is a solution that satisfies the constraints iff there is one if we ignore the constraints (the reason being that the constraints together form a finite lower bound while there is an infinite number of solutions). If there is a solution it will be a unique solution modulo $\lcm(a_1,a_2,\ldots,a_k)$ corresponding to an infinite number of solutions if expanded. From these solutions we just have to keep those that are not less than $t_{min}$ (in case we want to find the actual solutions to the system). 

\section{Random Temporal Graphs}
\label{sec:random}

Another model of temporal graphs with succinct representation, is the model of random temporal graphs. Consider the case in which each edge (of an underlying clique) just picks independently and uniformly at random \emph{a single time-label} from $[r]=\{1,2,\ldots,r\}$. So it gets label $t\in[r]$ with probability $p=r^{-1}$.

We first calculate the probability that given a specific path $(u_1,u_2,\ldots,u_{k+1})$ of length $k$ a journey appears on this path. We begin with the directed case. First, let us obtain a weak but elegant upper bound. Partition $[r]$ into $R_1=\{1,\ldots,\lfloor r/2\rfloor\}$ and $R_2=\{\lfloor r/2\rfloor+1,\ldots,r\}$. Clearly, $\P(journey)\leq \P(\text{no } R_2R_1 \text{ occurs})$ as any journey assignment cannot have two consecutive selections s.t. the first one is from $R_2$ and the second from $R_1$. So, it suffices to calculate $\P(\text{no } R_2R_1 \text{ occurs})$. Notice that the assignments in which no $R_2R_1$ occurs are of the form $(R_1)^{i}(R_2)^{j}$ for $i+j=k$, e.g. $R_1R_1R_2R_2R_2$ and there are $k+1$ of them. In contrast, all possible assignments are $2^k$ corresponding to all possible ways to choose $k$ times with repetition from $\{R_1,R_2\}$. So, $\P(\text{no } R_2R_1 \text{ occurs})=k/2^k$ (as all assignments are equiprobable, with probability $2^{-k}$) and we conclude that $\P(journey)\leq k/2^k$, which, interestingly, is independent of $r$; e.g. for $k=6$ we get a probability of at most $0.09375$ for a journey of length $6$ to appear.

For any specific assignment of labels $t_1,t_2,\ldots,t_k$ of this path, where $t_i\in [r]$ ($[r]=\{1,2,\ldots,r\}$), the probability that this specific assignment occurs is simply $p^k$. So, all possible assignments are equiprobable and we get
\begin{align*}
\P(journey)&=\frac{\# \text{ strictly increasing assignments}}{\# \text{ all possible assignments}}=\frac{\binom{r}{k}}{r^k},
\end{align*}  
where $\binom{r}{k}$ follows from the fact that any strictly increasing assignment is just a unique selection of $k$ labels from the $r$ available and any such selection corresponds to a unique strictly increasing assignment. So, for example, for $k=2$ and $r=10$ we get a probability of $9/20$ which is a little smaller than $1/2$ as expected, due to the fact that there is an equal number of strictly increasing and strictly decreasing assignments but we also loose all remaining assignments which in this case are only the ties (that is, those for which $t_1=t_2$).

Now it is easy to compute the expected number of journeys of length $k$. Let $S$ be the set of all directed paths of length $k$ and let $Y_p$ be an indicator random variable which is 1 if a journey appears on a specific $p\in S$ and 0 otherwise. Let also $X_k$ be a random variable giving the number of journeys of length $k$. Clearly, $\E(X_k)=\E(\sum_{p\in S} Y_p)=\sum_{p\in S} \E(Y_p)=|S|\cdot\P($a journey appears on a specific path of length $k)=n(n-1)\cdots(n-k)\binom{r}{k}r^{-k}\geq (n-k)^k\binom{r}{k}r^{-k}$. Now, if we set $n\geq r/\binom{r}{k}^{(1/k)}+k$, we get $E(X)\geq 1$. A simpler, but weaker, formula can be obtained by requiring $n\geq r+k$. In this case, we get $E(X)\geq \binom{r}{k}$. So, for example, a long journey of size $k=n/2$ that uses all available labels is expected to appear provided that $n\geq 2r$ (to see this, simply set $k=r$).

We will now try to obtain bounds on the probability that a journey of length $k$ appears on a random temporal graph. Let us begin from a simple case, namely the one in which $k=4$, that is, we want to calculate the probability that a journey of length 4 appears. Let the r.v. $X$ be the number of journeys of length 4 and let $X_p$ be an indicator for path $p\in S$, where $S$ is the set of all paths of length 4. Denote $n(n-1)\cdots(n-k)$ by $(n)_{k+1}$ First note that $E(X)=(n)_{5}\binom{r}{4}r^{-4}= \Theta(n^5)$ and clearly goes to $\infty$ for every $r$. However, we cannot yet conclude that $\P(4-journey)$ is also large. To show this we shall apply the second moment method. We will make use of Chebyshev's inequality $\P(X=0)\leq \Var(X)/[\E(X)]^2$ and of the following well-known theorem:

\begin{theorem} \label{the:secmom}
Suppose $X=\sum_{i=1}^n X_i$, where $X_i$ is an indicator for event $A_i$. Then,
\begin{equation*}
\Var(X)\leq \E(X)+\sum_i \P[A_i]\underbrace{\sum_{j:j\sim i}\P(A_j\mid A_i)}_{\Delta_i},
\end{equation*}
where $i\sim j$ denotes that $i$ depends on $j$. Moreover, if $\Delta_i\leq \Delta$ for all $i$, then
\begin{equation*}
\Var(X)\leq \E(X)(1+\Delta).
\end{equation*}
\end{theorem}

So, in our case, we need to estimate $\Delta_p=\sum_{p^\prime\sim p} \P(A_{p^\prime}\mid A_p)$. If we show that $\Delta_p\leq \Delta$ for all $p\in S$ then we will have that $\Var(X)\leq \E(X)(1+\Delta)$. If we additionally manage to show that $\Delta/E(X)=o(1)$, then $\Delta=o(\E(X))$ which tells us that $\Var(X)=o([\E(X)]^2)$. Putting this back to Chebyshev's inequality we get that $\P(X=0)=o(1)$ as needed. 

So, let us try to bound $\Delta_p$ appropriately. Clearly, $p^\prime$ cannot be a journey if it visits some edges of $p$ in inverse order (than the one they have on $p$). Intuitively, the two paths must have the same orientation. We distinguish cases based on the number of edges shared by the two paths. First of all, note that if $p^\prime$ and $p$ have precisely $i$ edges in common then $\P(A_{p^\prime}\mid A_p)\leq\binom{r}{k-i}/r^{k-i}$ which becomes $\binom{r}{4-i}/r^{4-i}$ in our case. The reason is that the $k-i$ edges of $p^\prime$ that are not shared with $p$ must at least obtain an increasing labeling. If we also had taken into account that that labeling should be consistent to the labels of the shared edges then this would decrease the probability. So we just use an upper bound which is sufficient for our purposes. 

\emph{Case 1: 1 shared edge.} If a single edge is shared then there are $k\binom{n-k+1}{k-1}(k-1)!4=16\cdot 3!\binom{n-5}{3}$ different paths $p^\prime$ achieving this as there are $k$ ways to choose the shared edge, $\binom{n-k+1}{k-1}$ to choose the missing nodes (nodes of $p^\prime$ not shared with $p$), $(k-1)!$ ways to order those nodes, and, in this particular example, 4 ways to arrange the nodes w.r.t. the shared edge. In particular, we can put all nodes before the shared edge, all nodes after, 2 nodes before and 1 node after, or 1 node before and 2 nodes after. We conclude that the probability that $\sum_{|p^\prime\cap p|=1} \P(A_{p^\prime}\mid A_p)\leq 16\cdot 3!\binom{n-5}{3}\binom{r}{3}/r^{3}=O(n^{3})$.

\emph{Case 2: 2 shared edges.} In this case, we can have all possible $\binom{k}{2}=\binom{4}{2}$ 2-sharings. Let us denote by $e_1,e_2,e_3,e_4$ the edges of $p$. For the sharings $(e_1,e_2)$, $(e_2,e_3)$, and $(e_3,e_4)$ we get in total $3\binom{n-k-1}{k-2}(k-2)!4=24\binom{n-5}{2}$ paths. For $(e_1,e_3)$, $(e_2,e_4)$ we get $2(n-k-1)=2(n-5)$. For $(e_1,e_4)$ we get $(n-5)$ in case we connect the 2 edges by an intermediate node (i.e. go from the head of $e_1$ to some $u$ not in $p$ and then form $u$ to the tail of $e_4$) and $2(n-5)$ in case we connect $e_1$ directly to $e_4$ and use an external node either before or after, so in total $3(n-5)$ paths. Putting these all together we get $\sum_{|p^\prime\cap p|=2} \P(A_{p^\prime}\mid A_p)\leq[24\binom{n-5}{2}+5(n-5)]\binom{r}{2}/r^{2}=O(n^2)$.

\emph{Case 3: 3 shared edges.} Here there are just 2 choices for the 3 shared edges, namely $(e_1,e_2,e_3)$ and $(e_2,e_3,e_4)$, the reason being that if the edges are not consecutive then a fourth edge must be necessarily shared and the 2 paths would coincide. As there are $(n-k-1)$ ways to choose the missing node and 2 ways to arrange that node we get $2(n-k-1)2=4(n-5)$ and consequently $\sum_{|p^\prime\cap p|=3} \P(A_{p^\prime}\mid A_p)\leq 4(n-5)\binom{r}{1}/r^{1}=O(n)$.

So, we have $\Delta_p\leq \Delta=O(n^3)$ and $\Delta/E(X)=O(n^3)/\Theta(n^5)=o(1)$ which applied to Theorem \ref{the:secmom} gives $\Var(X)\leq \E(X)(1+\Delta)=o([\E(X)]^2)$ and this in turn applied to Chebyshev's inequality gives the desired $\P(X=0)\leq \Var(X)/[\E(X)]^2=o(1)$. We conclude that:

\begin{theorem}
For all $r\geq 4$, almost all random temporal graphs contain a journey of length 4.
\end{theorem}

Now let us turn back to our initial $(n)_{k+1}\binom{r}{k}r^{-k}$ formula of $\E(X)$ (which holds for all $k$). This gives $\E(X)\geq (n)_{k+1}/k^k$, which, for all $k=o(n)$ and all $r\geq k$, goes to $\infty$ as $n$ grows. We will now try to generalize the ideas developed in the $k=4$ case to show that for any not too large $k$ almost all random temporal graphs contain a journey of length $k$. Take again a path $p$ of length $k$ and another path $p^\prime$ of length $k$ that shares $i$ edges with $p$. We will count rather crudely but in a sufficient way for our purposes. As again the shared edges can be uniquely oriented in the order they appear on $p$, there are at most $\binom{k}{i}$ ways to choose the shared edges (at most because some selections force more than $i$ sharings to occur). Counting the tail of the first edge and the head of every edge, these $i$ edges occupy at least $i+1$ nodes, so at most $k+1-i-1=k-i$ nodes are missing from $p^\prime$ and thus there are at most $\binom{n-k-1}{k-i}$ ways to choose those nodes. Moreover there are at most $(k-i)!$ ways to permute them on $p^\prime$. Finally, we have to place those nodes relative to the $i$ shared edges. In the worst case, the $i$ edges define $i+1$ slots that can be occupied by the nodes in $\binom{k-i+(i+1)-1}{(i+1)-1}=\binom{k}{i}$ ways. In total, we have $N=\binom{k}{i}^2\binom{n-k-1}{k-i}(k-i)!\binom{k}{i}$ different paths and the corresponding probability is 
\begin{align*}
\sum_{|p^\prime\cap p|=i}\P(A_{p^\prime}\mid A_p)&\leq N\binom{r}{k-i}/r^{k-i}\leq \binom{k}{i}^2\binom{n-k-1}{k-i}\\
&\leq \binom{k^2}{i}\binom{n-k-1}{k-i}.  
\end{align*}
So we have that
\begin{align*}
\Delta_p &=\sum_{i=1}^{k-1}\sum_{|p^\prime\cap p|=i}\P(A_{p^\prime}\mid A_p)\leq \sum_{i=0}^k \binom{k^2}{i}\binom{n-k-1}{k-i}\\
&=\binom{n+k^2-k-1}{k}\leq \binom{n+k^2}{k}=\Delta.
\end{align*}
The first equality follows from the Chu-Vandermonde identity $\sum_{i=0}^k\binom{m}{i}\binom{z}{k-i}=\binom{m+z}{k}$ by setting $z=n-k-1$ and $m=k^2$ as needed in our case.

Thus, we have $\Delta=\binom{n+k^2}{k}$ and for $k^2=o(n)$ we have $\Delta\sim (n)_k/k!$. At the same time we have $\E(X)=(n)_{k+1}\binom{r}{k}/r^k\sim (n)_{k+1}/k!$ (for large $r$), thus $\Delta/\E(X)\sim (n)_k/(n)_{k+1}=o(1)$ as needed. So we have $\Var(X)=o([\E(X)]^2)$ and we again get that $\P(X=0)\leq \Var(X)/[\E(X)]^2=o(1)$. Captured in a theorem:

\begin{theorem}
For all $k=o(\sqrt{n})$ and all $r=\Omega(n)$, almost all random temporal graphs contain a journey of length $k$.
\end{theorem}
However, there seems to be some room for improvements if one counts more carefully.

Now take any two nodes $s$ and $t$ in $V$. We want to estimate the arrival time of a foremost journey from $s$ to $t$. Let $X$ be the random variable of the arrival time of the foremost $s$-$t$ journey. Let us focus on $\P(X\leq 2)$. Denote by $l(u,v)$ the label chosen by edge $(u,v)$. Given a specific node $u\in V\bs\{s,t\}$ we have that $\P(l(s,u)\neq 1 \text{ or } l(u,t)\neq 2)=1-\P(l(s,u)= 1 \text{ and } l(u,t)= 2)=1-r^{-2}$. Thus, $\P(\forall u\in V\bs\{s,t\}: l(s,u)\neq 1 \text{ or } l(u,t)\neq 2)=(1-r^{-2})^{n-2}$  We have:
\begin{align*}
\P(X\leq 2)&=1-\P(X>2)\\
	   &=1-\P(l(s,t)\notin\{1,2\})\P(\forall u\in V\bs\{s,t\}: l(s,u)\neq 1 \text{ or } l(u,t)\neq 2)\\
	   &=1-\frac{r-2}{r}(1-r^{-2})^{n-2}\\
	   &\geq 1-(1-r^{-2})^{n-2}\\
	   &\geq 1-e^{-(n-2)/r^2}, \text{ for } n\geq 2 \text{ and } r>\sqrt{n-1}.
\end{align*}

So, even if $r=\Theta(\sqrt{n})$ we have that $\P(X\leq 2)\ra 1-1/e^c$ (for some constant $c\leq 1$) as $n$ goes to infinity, so we have a constant probability of arriving by time 2 at $t$. Clearly, for smaller values of $r$ (smaller w.r.t. $n$) we get even better chances of arriving early. For another example, let $n=10^4$ and $r=\sqrt{n}/\log n=25$. As $\P(X\leq 2)$ is almost equal to $1-(1-r^{-2})^{n-2}$ we get that it is almost equal to 1 in this particular case. For even greater $r$, e.g. $r=\sqrt{n}=100$, we still go very close to 1.

The following proposition gives a bound on the temporal diameter of undirected random temporal graphs, by exploiting well known results of the Erd\" os-Renyi ($\cg(n,p)$) model (cf. \cite{Bol01}).

\begin{proposition}
Almost no temporal graph has temporal diameter less than $[(\ln n +c+o(1))/n]r$.
\end{proposition}

To see this, observe that if $k<[(\ln n +c+o(1))/n]r$ then $p=k/r<(\ln n +c+o(1))/n$. Consider now the temporal subgraph consisting only of the first $k$ labels $[k]=\{1,2,\ldots,k\}$. By the connectivity threshold of the static $\cg(n,p)$ model this subgraph is almost surely disconnected implying that almost surely the temporal diameter is greater than $k$.

So, for example, if $r=O(n)$ almost no temporal graph has temporal diameter $o(\log n)$. Note, however, that the above argument is not sufficient to show that a.e. temporal graph has temporal diameter at least $[(\ln n +c+o(1))/n]r$. Though it shows that in a.e. graph the subgraph consisting of the labels $[k]$, for $k\geq\lceil r(\ln n +c+o(1))/n\rceil$ is connected, it does not tell us whether that connectivity also implies temporal connectivity (that is, the existence of journeys).

We should also mention that \cite{AGMS14} studied the temporal diameter of the directed random temporal graph model for the case of $r=n$, and proved that it is $\Theta(\log n)$ w. and in expectation. In fact, they showed that information dissemination is very fast w.h.p. even in this hostile network with regard to availability. Moreover, they showed that the temporal diameter of the clique is crucially affected by the clique's lifetime, $\alpha$, e.g., when $\alpha$ is asymptotically larger than the number of vertices, $n$, then the temporal diameter must be $\Omega(\frac{\alpha}{n}\log n)$. They also defined the \emph{Price of Randomness} metric in order to capture the cost to pay per link and guarantee temporal reachability of all node-pairs by local random available times w.h.p..

The idea of \cite{AGMS14} to establish that the temporal diameter is $O(\log n)$ is as follows. Given an instance of such a random temporal clique, the authors pick any source node $s$ and any sink node $t$ and present an algorithm trying to construct a journey from $s$ arriving at $t$ at most by time $O(\log n)$. The algorithm expands two fronts, one beginning from $s$ and moving forward (in fact, an out-tree rooted at $s$) and one from $t$ moving backward (an in-tree rooted at $t$). Beginning from $s$, all neighbors that can be reached in one step in the interval $(0,c_1\log n]$, are visited. Next the front moves on to all neighbors of the previous front that can be reached in one step in the interval $(c_1\log n,c_1\log n +c_2]$. The process continues in the same way, every time replacing the current front by all its neighbors that can be reached in the next $c_2$ steps. A similar backward process is executed from $t$. These processes are executed for $d=\Theta(\log n)$ steps resulting in the final front of $s$ and the final front of $t$. Note that the front of $t$ begins from the interval $(2c_1\log n+(2d-1)c_2,2c_1\log n+2dc_2]$ and every time subtracts a $c_2$. Finally, the algorithm tries to find an edge from the final front of $s$ to the final front of $t$ with the appropriate label in order to connect the journey from $s$ to the journey to $t$ in a time-respecting way and obtain the desired $s$-$t$ journey of duration $\Theta(\log n)$ (determined by the interval of the first front of $t$, and in particular by $2c_1\log n+2dc_2$). Via probabilistic analysis it can be proved that, with probability at least $1-1/n^3$, the final front of $s$ consists of $\Theta(\sqrt{n})$ nodes and that the same holds for the front of $t$. Moreover, it can be proved that again with probability at least $1-1/n^3$ the desired edge for the final front of $s$ to the final front of $t$ exists, and thus we can conclude that there is a probability of at least $1-3/n^3$ of getting from $s$ to $t$ by a journey arriving at most by time $\Theta(\log n)$. Finally, it suffices to observe that the probability that there exists a pair of nodes $s,t\in V$ for which the algorithm fails is less than $n^2(3/n^3)=3/n$, thus with probability at least $(1-3/n)$ the temporal diameter is $O(\log n)$, as required.

\newcommand{\etalchar}[1]{$^{#1}$}

\end{document}